\documentclass[12pt]{article}
\usepackage[utf8]{inputenc}
\usepackage[british]{babel}
\usepackage{cmap}
\usepackage{lmodern}
\usepackage[T1]{fontenc}

\usepackage{amssymb, amsmath, amsthm}
\usepackage[a4paper,top=25mm,bottom=25mm,left=25mm,right=25mm]{geometry}
\usepackage{ragged2e}

\usepackage{authblk} 
\usepackage{pifont}
\usepackage{graphicx}
\usepackage[dvipsnames,svgnames,table]{xcolor}
\usepackage[figuresright]{rotating}
\usepackage{xtab} 
\usepackage{longtable} 
\usepackage{multirow}
\usepackage{footnote}
\usepackage[stable]{footmisc}
\usepackage{chngpage} 
\usepackage{pdflscape} 
\usepackage[nottoc,notlot,notlof]{tocbibind} 

\usepackage{pgfplots}
\pgfplotsset{every tick label/.append style={font=\footnotesize}}
\pgfplotsset{compat=1.18}
\usepackage{setspace}

\usepackage{array}
\newcolumntype{K}[1]{>{\centering\arraybackslash$}p{#1}<{$}}

\makesavenoteenv{tabular}
\usepackage{tabularx}
\usepackage{booktabs}
\usepackage{threeparttable}
\usepackage[referable]{threeparttablex} 
\newcolumntype{R}{>{\raggedleft\arraybackslash}X}
\newcolumntype{L}{>{\raggedright\arraybackslash}X}
\newcolumntype{C}{>{\centering\arraybackslash}X}
\newcolumntype{A}{>{\columncolor{gray!25}}C}
\newcolumntype{a}{>{\columncolor{gray!25}}c}

\newlength{\tablen}

\usepackage{dcolumn} 
\newcolumntype{.}{D{.}{.}{-1}}

\usepackage{tikz}
\usetikzlibrary{arrows, calc, matrix, patterns, positioning, trees}
\usepackage[semicolon]{natbib} 
\usepackage[hyphens]{xurl}
\usepackage[nopatch=footnote]{microtype}
\usepackage[justification=centering]{caption} 

\usepackage[labelformat=simple]{subcaption}

\DeclareCaptionLabelFormat{parenthesis}{(#2)}
\makeatletter
\renewcommand\p@subfigure{\arabic{figure}.}
\makeatother

\DeclareCaptionLabelFormat{parenthesis}{(#2)}
\makeatletter
\renewcommand\p@subtable{\arabic{table}.}
\makeatother

\usepackage{enumitem}

\setlist[itemize]{leftmargin=2.5\parindent}
\setlist[enumerate]{leftmargin=2.5\parindent}

\usepackage{hyperref} 
\hypersetup{
  colorlinks   = true,    		
  urlcolor     = blue,    		
  linkcolor    = blue,    		
  citecolor    = ForestGreen	
}

\def\addlegendimage{\csname pgfplots@addlegendimage\endcsname}

\theoremstyle{plain}

\theoremstyle{definition}

\newtheorem{example}{Example}

\theoremstyle{remark}

\makeatletter
\let\@fnsymbol\@alph
\makeatother

\def\keywords{\vspace{.5em} 
{\noindent \textit{Keywords}: }}

\def\AMS{\vspace{.5em} 
{\noindent \textbf{\emph{MSC} class}: }}

\def\JEL{\vspace{.5em} 
{\noindent \textbf{\emph{JEL} classification number}: }}

\title{Ranking-based competitive balance \\ measures in Formula One}
\author{
\href{https://sites.google.com/view/doragretapetroczy}{ D\'ora Gr\'eta Petr\'oczy}\thanks{~E-mail: \emph{apetroczy@metropolitan.hu} \newline Budapest Metropolitan University, Budapest, Hungary}
$\qquad \qquad$
\href{https://sites.google.com/view/laszlocsato}{L\'aszl\'o Csat\'o}\thanks{~E-mail: \emph{laszlo.csato@sztaki.hun-ren.hu}
\newline Institute for Computer Science and Control (SZTAKI), Hungarian Research Network (HUN-REN), Laboratory on Engineering and Management Intelligence, Research Group of Operations Research and Decision Systems, Budapest, Hungary
\newline Corvinus University of Budapest (BCE), Institute of Operations and Decision Sciences, Department of Operations Research and Actuarial Sciences, Budapest, Hungary}
}

\date{\today}

\begin{document}

\maketitle

\begin{abstract}
\noindent
Competitiveness in racing sports can be measured by comparing the start and finish rankings within races, as well as the start and finish rankings across races in a season. Since the importance of position changes is non-uniform and variance at the top of the ranking is more interesting than at the bottom of the ranking, we propose using weighted distances for this purpose. Therefore, two weighting schemes are applied and compared to the standard Kem\'eny distance to analyse the Formula One between 1950 and 2024. The evolution of competitive balance is unexpectedly robust to the weights, but competition is more intense if one focuses on the top positions. Competitive balance has been more unfavourable in the last two decades than ever before. Statistical tests uncover three structural breaks in competitive balance that are closely related to regulatory changes, highlighting the role of decision-makers in the evolution of competitiveness.

\keywords{competitive balance; Formula One; ranking; sports economics; weighted distance}

\AMS{62P20, 91-10, 91B14}

\JEL{D63, L83, Z20}
\end{abstract}

\clearpage

\section{Introduction}

The seminal paper \citet{Szymanski2003} raises the following question as a prominent example of a design issue in sports: \emph{How evenly balanced should the competing teams be in the NASCAR or Formula One championships?}
The answer to this problem calls first for an appropriate measure of competitive balance. 
\citet{PeetersWesselbaum2023} have recently suggested quantifying competitive balance in racing sports based on the idea that each race establishes a ranking of the contestants, and the variability of this ranking over the season is directly related to outcome uncertainty. However, using Kem\'eny distance \citep{Kemeny1959, KemenySnell1962}---the number of pairs of alternatives ranked in an opposite order---to measure the dissimilarity between two rankings is debatable, as its value does \emph{not} depend on where the switches occur. Consequently, the distance between two race results is the same if the first two or the last two drivers are exchanged.

Contrarily, it is reasonable to assume that the ranking places need to be distinguished according to their significance for fans \citep{ManasisNtzoufras2014}. In sports, the race winner is usually more important than the runner-up, the second place is more important than the third, and so on. Indeed, even though Formula One has used several point scoring systems in its history, the differences between the value of subsequent positions have always been monotonically decreasing \citep{Haigh2009, Csato2023b}.

Therefore, instead of the Kem\'eny distance, we will use a class of semi-metrics introduced by \citet{Can2014}, which allows quantifying disagreements according to their positions in the ranking. This measure generalises the Kem\'eny metric on strict preferences. It satisfies the triangle inequality and can be calculated relatively easily if the associated weights are monotonically decreasing from the top to the bottom of the ranking, as assumed by us.

Our main contribution to the extensive literature on competitive balance resides in suggesting novel indicators based on the weighted distance of \citet{Can2014}. The parametric method can be used to quantify the (dis)similarity of two strict rankings. The only restriction on the weights assigned is that they should be monotonically decreasing (or increasing, but this is not relevant in the context of sports). The exact weights may be determined by questionnaires or empirical estimations; alternatively, a wide set of weights is worth considering to explore the robustness of the findings.

We apply these competitive balance measures to Formula One by comparing start and finish rankings within races, as well as start rankings and finish rankings in subsequent races of a season.
Since the race results rank the drivers rather than the teams, our focus is on driver-level competitive balance in order to avoid the numerous problems caused by deriving team-corrected rankings.
The paper contains two further novelties in measuring competitive balance in Formula One. First, we use the Bai--Perron test to automatically detect and date structural breaks with the aim of connecting them to regulatory changes. Second, we consider the longest possible dataset, covering all seasons from 1950 to 2024.

The structure of the study is as follows. Section~\ref{Sec2} provides a concise overview of related papers, focusing on competitive balance and its effects in Formula One.
Section~\ref{Sec3} introduces the methodology and data used to conduct the numerical analysis in Section~\ref{Sec4}. Section~\ref{Sec5} ends with concluding remarks.

\section{Literature review} \label{Sec2}

Formula One is often considered as the pinnacle of motorsport. Unsurprisingly, its competitive balance has been the subject of several academic papers.
\citet{MastromarcoRunkel2009} is the first study connecting rule changes to competitive balance in Formula One. Based on the seasons from 1950 to 2005, they find evidence that rule changes usually follow unbalanced seasons and significantly improve competitive balance in the subsequent years.
\citet{KrauskopfLangenBunger2010} introduce the `superstar' effect: attractiveness is increased not just by better competitive balance, but also by the duel at the top.
\citet{JuddeBoothBrooks2013} examine competitive balance between 1950 and 2010 using indicators such as the championship decision point, the number of lead changes per race, and concentration indices based on wins and points. Rule changes are shown to significantly increase championship uncertainty by keeping title races alive longer, while they have a limited impact on race-level uncertainty or long-term dominance. 
\citet{BudzinskiFeddersen2020} broaden the analysis by distinguishing within-race, within-season, and inter-season competitive balance and considering several indicators, including lead changes, Gini coefficients, leading laps, leading distance, and rank-correlation measures. Conclusions about competitive balance are found to strongly depend on the measurement approach, although several indicators suggest a modest long-run improvement in competitiveness, with signs of renewed dominance since 2010.
\citet{LeePrinceZeager2026} use Gini coefficients and Lorenz curves to compare the inequality in lap times in the 2021 and 2022 seasons. Quantile regressions reveal that lap times decreased for the slowest drivers and increased for faster drivers in 2022, implying a reduced lap-time dispersion.

Analogous to team sports \citep{Groot2008, CsatoPetroczy2025a}, ranking-based approaches have recently been suggested to measure competitive balance in Formula One. \citet{PeetersWesselbaum2023} use a method based on Kem\'eny distance \citep{Kemeny1959}, which captures the number of positional changes---interpreted as the minimal number of overtakes---required to transform one race ranking into another. This approach improves on traditional indices (e.g.\ Gini, Herfindahl--Hirschman index, standard deviation) by incorporating the full distribution of positions rather than focusing on winners or points totals. Similarly, \citet{Pedroche2024} applies a ranking-based normalised strength index derived from Kendall-type measures to evaluate competitive balance across seasons. 

Using race-by-race rankings, \citet{Pedroche2024} finds relatively stable levels of competitiveness between 2012 and 2022, with similar patterns across the championships of drivers and constructors.
In contrast, \citet{PeetersWesselbaum2023} identify temporal variation linked to regulatory changes, including a decline in competitiveness during the mid-2000s to early 2010s, associated with rule changes such as engine restrictions. Races have become increasingly predictable over time, particularly due to reduced mechanical failures and more homogeneous strategic conditions. 


The second major strand of literature investigates the economic implications of competitive balance, particularly its impact on spectator demand. Classical theory predicts that higher uncertainty increases fan interest; however, empirical evidence from Formula One remains mixed and often contradictory.
According to \citet{SchreyerTorgler2018}, TV demand is shaped by race outcome uncertainty, proxied by performance differences among drivers with the best qualifying performances.
Based on Google Trends data, a third of search interest for the Formula One world drivers’ championship is lost when pennant-race uncertainty is removed \citep{Garcia-del-BarrioReade2022}.
On the other hand, \citet{GasparettoOrlovaVernikovskiy2024} find a non-linear (inverted U-shaped) relationship between attendance and uncertainty; hence, both overly predictable and excessively random outcomes reduce demand. 
\citet{BaeckerAnsariSchreyer2024} demonstrate that the effect of outcome uncertainty is context-dependent, varying across broadcast channels, audiences, and measurement methods, and is often weaker than other factors such as race timing or accessibility.
\citet{FahyButlerButler2026} do not find any evidence to support the uncertainty of outcome hypothesis: the audience in the United States may even prefer less competitive races, with higher viewership associated with greater differences in betting odds, that is, more predictable outcomes.

To summarise, competitive balance in Formula One is moderate, rule-sensitive, and not necessarily positively linked to demand. While recent advances in measurement---particularly ranking-based methods---have improved our understanding of competitiveness, empirical evidence on its effects is far from conclusive.

\section{Data and methodology} \label{Sec3}

Section~\ref{Sec31} presents the underlying data, and Section~\ref{Sec32} describes the proposed method to measure competitive balance in Formula One. Finally, Section~\ref{Sec33} summarises the background of identifying structural breaks.

\subsection{Data} \label{Sec31}

We have downloaded the \href{https://www.kaggle.com/datasets/rohanrao/formula-1-world-championship-1950-2020/}{Formula 1 World Championship (1950--2024)} database from Kaggle, compiled from the ERGAST API.
This contains the start and finish positions of all the 1125 races in the 75 seasons between 1950 and 2024 (15 races per year on average). The number of drivers and races each year is given in Table~\ref{Table_A1} in the Appendix. Note the declining trend in the number of drivers and the consistent increase in the number of races: in recent years, the two values are roughly equal around 20, while there were usually at most 10 races in the 1950s with more than 80 drivers.

According to our definition, if a driver does not qualify or starts the race from the pit lane, he is assigned to the last starting position. For example, Valtteri Bottas qualified as the 17th in the 2024 Canadian Grand Prix, but was required to start the race from the pit lane; thus, we assign the 19th position to him.
In the case of more than one driver starting from the pit lane, the final race result is used as a tie-breaking rule in the last positions. 

The official methodology is followed for the final ranking; retired drivers are ranked according to their number of laps finished. For instance, the last driver who finished the 2024 Canadian Grand Prix was the 15th Zhou Guanyu. Five drivers retired in the race due to accident, collision, or engine failure. Among them, Carlos Sainz Jr.\ was the 16th by performing 52 laps, and Logan Sargeant was the 20th since he retired after 23 laps.

\subsection{Weighted distance between rankings} \label{Sec32}

\citet{PeetersWesselbaum2023} use the Kem\'eny distance \citep{Kemeny1959, KemenySnell1962} to measure competitive balance in Formula One. This is simply the number of pairwise disagreements between two rankings. However, although the minimum number of overtakes that should have occurred during a race is indeed relevant, it neglects the position-dependent importance of the positional changes. Therefore, as discussed in the Introduction, we use the weighted distance proposed by \citet{Can2014}.


The Kem\'eny distance of the reference and observed rankings counts the number of \emph{adjacent swaps} in an arbitrary order. In contrast, the weighted distance of \citet{Can2014} should be computed by the winners’ decomposition if the weights are monotonically decreasing in order to obtain a measure satisfying the triangle inequality. Starting from an observed ranking, the winners’ decomposition permutes first the winner of the reference ranking to the top, then the runner-up to the second place, and so on, as illustrated below.

\begin{example} \label{Examp1}
Consider four items $A$, $B$, $C$, and $D$.
The observed ranking is $C > A > D > B$, and the reference ranking is $A > B > C > D$. We want to create the reference ranking from the observed ranking.

In the first step, $A$ should be moved to the first position, which can be achieved by one swap between positions $1$ and $2$:
\[
C > A > D > B \quad \to \quad A > C > D > B
\]
In the second step, $B$ needs to be moved to the second position, which can be done by two swaps, one between positions $3$ and $4$, and the next between positions $2$ and $3$:
\[
A > C > D > B 
\quad \to \quad
A > C > B > D 
\quad \to \quad
A > B > C > D
\]
The reference ranking is created, no further swap is necessary.

The total number of adjacent swaps is $d_K = 3$: one between positions 1 and 2, one between positions 2 and 3, and one between positions 3 and 4.
\end{example}

For $n$ items, the maximum number of pairwise disagreements is $n(n-1)/2$.
Hence, the normalised Kem\'eny distance is
\[
\tilde{d}_K = \frac{2d_K}{n(n-1)},
\]
which equals 0.5 in Example~\ref{Examp1}.

In the case of Formula One, it is reasonable to distinguish swaps between different positions. 
Therefore, we use the weighted distance of \citet{Can2014}:
\[
d_C^{(\mathbf{w})} = \sum_{k=1}^{n-1} w_k s_k,
\]
where $s_k$ denotes the number of swaps between positions $k$ and $k+1$, and $\mathbf{w} = \left[ w_k \right]$ is a nonnegative, monotonically decreasing $(n-1)$-dimensional weight vector.
Note that $d_C^{(\mathbf{w})} = d_K$ if $w_k = 1$ for all $1 \leq k \leq n-1$.

The maximum of $d_C^{(\mathbf{w})}$ is achieved when two opposite rankings are compared, which implies $k$ swaps between positions $k$ and $k+1$.
Hence, the normalised weighted distance is
\[
\tilde{d}_C^{(\mathbf{w})} = \frac{\sum_{k=1}^{n-1} w_k s_k}{\sum_{k=1}^{n-1} w_k k}.
\]

We consider three sets of weights:
\begin{itemize}
\item
Standard Kem\'eny distance: $w_k = 1$ for all $1 \leq k \leq n-1$;
\item
The top-heavy inverse weighting \citep{Csato2017a}: $w_k = 1/k$ for all $1 \leq k \leq n-1$;
\item
The inverse square root weighting \citep{Ausloos2024a}: $w_k= \sqrt{1/k}$ for all $1 \leq k \leq n-1$.
\end{itemize}
In the case of the top-heavy weighting $w_k=1/k$, the maximum is $d_{C,\max}^{(\mathbf{w})} = n-1$. Thus, $\tilde{d}_C^{(\mathbf{w})} = (1 + 1/2 + 1/3)/3 = 11/18 \approx 0.611$ in Example~\ref{Examp1}, and competitive balance is more favourable when changes in the top positions are judged to be more important.

The value of our competitiveness measure, the normalised weighted distance $\tilde d_C^{(\mathbf{w})}$, is between 0 and 1; a higher value means more variation and more intense competition.

Analogous to \citet{PeetersWesselbaum2023}, the following indicators of competitive balance are computed each year, consisting of $m$ races:
\begin{itemize}
\item 
Weighted distance of start--finish rankings: the average normalised distance between the rankings implied by start and finish positions in all races;
\item 
Weighted distance of start--start rankings: the average normalised distance between the rankings implied by start positions in all the possible $m(m-1)$ pairs of races (restricted to the set of drivers who took part in both races);
\item 
Weighted distance of finish--finish rankings: the average normalised distance between the rankings implied by finish positions in all the possible $m(m-1)$ pairs of races (restricted to the set of drivers who took part in both races).
\end{itemize}

\subsection{Identifying structural breaks} \label{Sec33}

The structural changes in our competitive balance measures are detected and dated by the Bai--Perron method \citep{BaiPerron1998, BaiPerron2003}.
This procedure has been used to identify significant changes in competitive balance for association football \citep{LeeFort2012}, baseball \citep{LeeFort2005}, and basketball \citep{MillsSalaga2015}, but, as far as we know, no previous study has applied it to analyse competitive balance in Formula One.

The method is designed to estimate unknown break dates in linear models by minimising the residual sum of squares over alternative segmentations of the sample. In our setting, the dependent variable is an annual competitive balance measure, and the model is specified as follows:
\[
\mathit{CB}_t = \alpha_j + \beta_j t + u_t, \qquad t = T_{j-1}+1,\ldots,T_j,
\]
where $\mathit{CB}_t$ denotes the competitive balance measure in season $t$, and and $\alpha_j$ and $\beta_j$ are regime-specific parameters in segment $j$. Consequently, the breakpoints indicate structural changes in the trend of competitive balance rather than merely a temporary fluctuation in one season. The breakpoints are estimated by the \texttt{breakpoints} function of the R package \texttt{strucchange} \citep{ZeileisLeischHornikKleiber2002}. We use the default setting of the package for the trimming parameter of minimum segment size, $h = 0.15$; each regime must contain at least 15\% of the observations. 
Crucially, the Bai--Perron procedure allows structural changes in competitive balance to be detected endogenously rather than imposing the number of breaks or their dates in advance.

\section{Results} \label{Sec4}

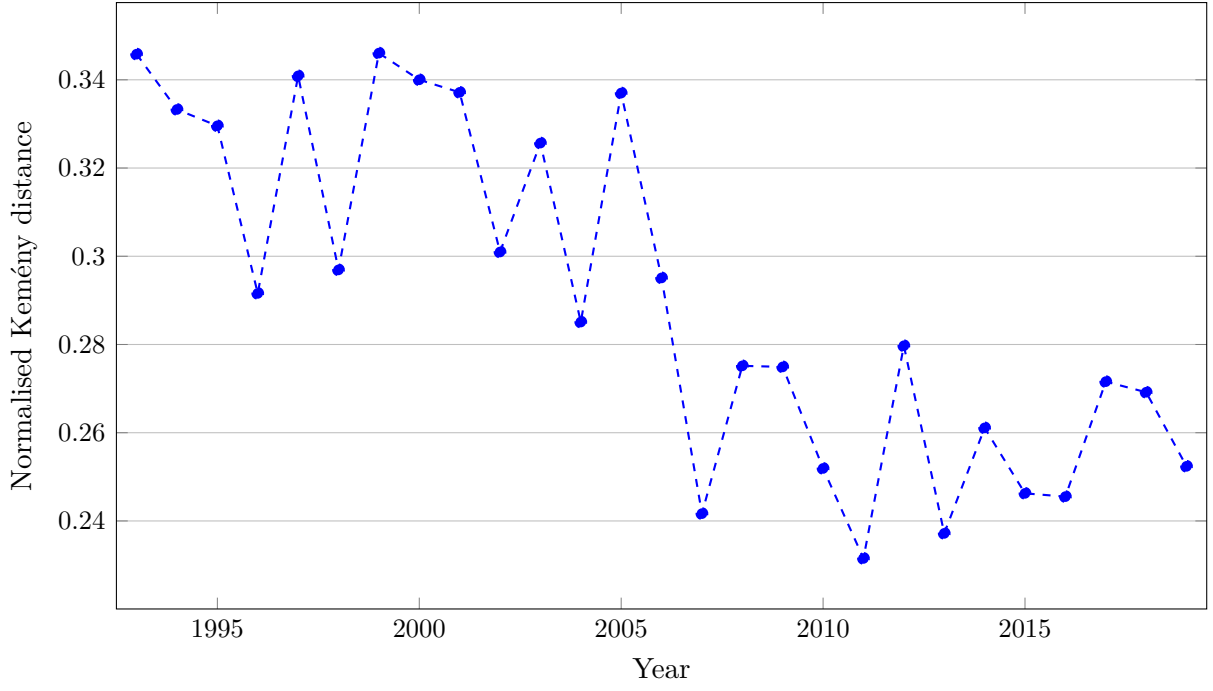
\begin{figure}[t!]
\centering

\begin{tikzpicture}
\begin{axis}[
width = \textwidth, 
height = 0.6\textwidth,
xlabel style = {align=center, font=\small},
xlabel = {Year},
ylabel style = {align=center, font=\small},
ylabel = {Normalised Kem\'eny distance},
xmin = 1992.5,
xmax = 2019.5,
ymajorgrids,
xtick = {1995,2000,2005,2010,2015},
scaled x ticks = false,
xticklabel style = {/pgf/number format/1000 sep=}
]

\addplot[color=blue, mark=*,dashed,thick] coordinates {
(1993,0.345813127090301)
(1994,0.333275844525844)
(1995,0.329569152075545)
(1996,0.291585497835497)
(1997,0.340889715570022)
(1998,0.296915584415584)
(1999,0.346049783549783)
(2000,0.340005092946269)
(2001,0.337153043035395)
(2002,0.300969000040207)
(2003,0.325657894736842)
(2004,0.285087719298245)
(2005,0.336995998768852)
(2006,0.295093795093795)
(2007,0.241660300483829)
(2008,0.275160122528543)
(2009,0.274922600619195)
(2010,0.251906941266209)
(2011,0.231502669717772)
(2012,0.279710144927536)
(2013,0.237183868762816)
(2014,0.261096497694183)
(2015,0.246254956276139)
(2016,0.245516388373531)
(2017,0.271578947368421)
(2018,0.269172932330827)
(2019,0.252380952380952)
};
\end{axis}
\end{tikzpicture}

\caption{Normalised Kem\'eny distance of start--finish rankings, 1993--2019}
\label{Fig1}
\end{figure}


First, we calculate the normalised Kem\'eny distance between the start and finish positions within all races in a season, and take their average. Thus, Figure~\ref{Fig1} is comparable to  \citet[Figure~2]{PeetersWesselbaum2023}, but they used different normalisation and focused on teams rather than on drivers. Nonetheless, the pattern is quite similar; only the trends around 2000 differ slightly.

\begin{figure}[t!]
\centering
\begin{tikzpicture}
\begin{axis}[
width = \textwidth, 
height = 0.6\textwidth,
xlabel style = {align=center, font=\small},
xlabel = {Year},
ylabel style = {align=center, font=\small},
ylabel = {Normalised Kem\'eny distance},
xmin = 1949.5,
xmax = 2024.5,
ymajorgrids,
xtick = {1950,1960,1970,1980,1990,2000,2010,2020}, 
scaled x ticks = false,
xticklabel style = {/pgf/number format/1000 sep=},
]

\addplot[color=blue,dashed,thick,mark=*,mark size=1pt] coordinates {
(1950,0.346785929344387)
(1951,0.324781858467442)
(1952,0.277069192864969)
(1953,0.307535714175066)
(1954,0.345924462838601)
(1955,0.349648030230443)
(1956,0.384570141187788)
(1957,0.296688804708263)
(1958,0.323457122281936)
(1959,0.287249928679871)
(1960,0.320389380362088)
(1961,0.238244871808843)
(1962,0.23277402289295)
(1963,0.248602841061986)
(1964,0.324994854900562)
(1965,0.26969010067012)
(1966,0.366416201426521)
(1967,0.409363547790657)
(1968,0.400525350378123)
(1969,0.384676133979539)
(1970,0.334210628710768)
(1971,0.330704441167845)
(1972,0.350501377053504)
(1973,0.363514188266548)
(1974,0.30934562929806)
(1975,0.346838498238155)
(1976,0.310423851372127)
(1977,0.296882267405565)
(1978,0.274693736987676)
(1979,0.30284968586118)
(1980,0.307677691766361)
(1981,0.244288068459369)
(1982,0.308015519889824)
(1983,0.352559265340874)
(1984,0.360066646316646)
(1985,0.368709565394348)
(1986,0.32318376068376)
(1987,0.357310885571755)
(1988,0.239423433444568)
(1989,0.187057756136703)
(1990,0.199110140725569)
(1991,0.217633222643896)
(1992,0.296738645950617)
(1993,0.345813127090301)
(1994,0.333275844525844)
(1995,0.329569152075545)
(1996,0.291585497835497)
(1997,0.340889715570022)
(1998,0.296915584415584)
(1999,0.346049783549783)
(2000,0.340005092946269)
(2001,0.337153043035395)
(2002,0.300969000040207)
(2003,0.325657894736842)
(2004,0.285087719298245)
(2005,0.336995998768852)
(2006,0.295093795093795)
(2007,0.241660300483829)
(2008,0.275160122528543)
(2009,0.274922600619195)
(2010,0.251906941266209)
(2011,0.231502669717772)
(2012,0.279710144927536)
(2013,0.237183868762816)
(2014,0.261096497694183)
(2015,0.246254956276139)
(2016,0.245516388373531)
(2017,0.271578947368421)
(2018,0.269172932330827)
(2019,0.252380952380952)
(2020,0.2656346749226)
(2021,0.233014354066985)
(2022,0.289712918660287)
(2023,0.270095693779904)
(2024,0.203435672514619)
};

\draw [red,thick,dotted] (1961,\pgfkeysvalueof{/pgfplots/ymin}) -- (1961,\pgfkeysvalueof{/pgfplots/ymax});
\draw [red,thick,dotted] (1966,\pgfkeysvalueof{/pgfplots/ymin}) -- (1966,\pgfkeysvalueof{/pgfplots/ymax});
\draw [red,thick,dotted] (1981,\pgfkeysvalueof{/pgfplots/ymin}) -- 
(1981,\pgfkeysvalueof{/pgfplots/ymax});
\draw [red,thick,dotted] (1989,\pgfkeysvalueof{/pgfplots/ymin}) -- (1989,\pgfkeysvalueof{/pgfplots/ymax});
\draw [red,thick,dotted] (1994,\pgfkeysvalueof{/pgfplots/ymin}) -- (1994,\pgfkeysvalueof{/pgfplots/ymax});
\draw [red,thick,dotted] (1998,\pgfkeysvalueof{/pgfplots/ymin}) -- (1998,\pgfkeysvalueof{/pgfplots/ymax});
\draw [red,thick,dotted] (2009,\pgfkeysvalueof{/pgfplots/ymin}) -- (2009,\pgfkeysvalueof{/pgfplots/ymax});
\draw [red,thick,dotted] (2011,\pgfkeysvalueof{/pgfplots/ymin}) -- (2011,\pgfkeysvalueof{/pgfplots/ymax});
\draw [red,thick,dotted] (2014,\pgfkeysvalueof{/pgfplots/ymin}) -- (2014,\pgfkeysvalueof{/pgfplots/ymax});
\draw [red,thick,dotted] (2022,\pgfkeysvalueof{/pgfplots/ymin}) -- (2022,\pgfkeysvalueof{/pgfplots/ymax});

\node[rotate=90, anchor=south, font=\footnotesize] at (axis cs:1961,0.9*\pgfkeysvalueof{/pgfplots/ymax}) {1.5L era};
\node[rotate=90, anchor=south, font=\footnotesize] at (axis cs:1966,0.9*\pgfkeysvalueof{/pgfplots/ymax}) {3.0L};
\node[rotate=90, anchor=south, font=\footnotesize] at (axis cs:1981,0.85*\pgfkeysvalueof{/pgfplots/ymax}) {sliding skirts, Concorde};
\node[rotate=90, anchor=south, font=\footnotesize] at (axis cs:1989,0.9*\pgfkeysvalueof{/pgfplots/ymax}) {Turbo ban};
\node[rotate=90, anchor=south, font=\footnotesize] at (axis cs:1994,0.9*\pgfkeysvalueof{/pgfplots/ymax}) {Safety reform};
\node[rotate=90, anchor=south, font=\footnotesize] at (axis cs:1998,0.9*\pgfkeysvalueof{/pgfplots/ymax}) {Narrow cars};
\node[rotate=90, anchor=south, font=\footnotesize] at (axis cs:2009,0.9*\pgfkeysvalueof{/pgfplots/ymax}) {Aero reform};
\node[rotate=90, anchor=south, font=\footnotesize] at (axis cs:2011,0.9*\pgfkeysvalueof{/pgfplots/ymax}) {DRS};
\node[rotate=90, anchor=south, font=\footnotesize] at (axis cs:2014,0.9*\pgfkeysvalueof{/pgfplots/ymax}) {Hybrid};
\node[rotate=90, anchor=south, font=\footnotesize] at (axis cs:2022,0.9*\pgfkeysvalueof{/pgfplots/ymax}) {Ground effect};

\end{axis}
\end{tikzpicture}

\caption{Normalised Kem\'eny distance of start--finish \\ rankings with major regulatory changes}
\label{Fig2}
\end{figure}
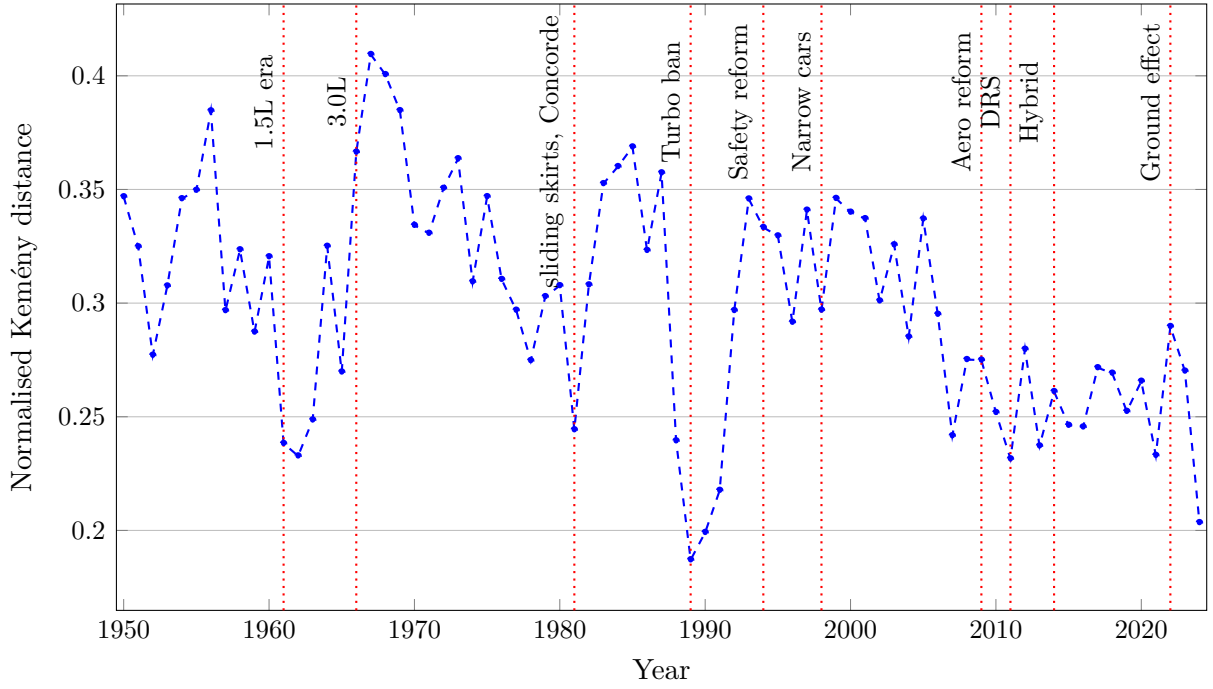


Figure~\ref{Fig2} shows the competitive balance values for the whole period from 1950 to 2024, highlighting the most important rule changes.
In 1961, the introduction of the 1.5-litre engine marked a fundamental shift toward lower-powered cars. This was followed in 1966 by the transition to the 3.0-litre era, leading to increased power and renewed competitive differentiation among teams.
Another major turning point occurred in 1989 with the ban on turbocharged engines, which eliminated a key source of performance disparity. The 1994 season implemented comprehensive safety reforms in response to fatal accidents (the San Marino Grand Prix saw the death of the three-time World Champion Ayrton Senna), including restrictions on electronic driver aids and aerodynamic features. In 1998, further regulatory reform resulted in narrower cars and grooved tyres. 

The 2009 aerodynamic overhaul, often associated with the emergence of Brawn GP, aimed to promote overtaking by simplifying airflow structures, thereby influencing race variability. In 2011, the introduction of the Drag Reduction System (DRS) provided a mechanical tool to facilitate overtaking, though its effectiveness depended on the underlying competitive parity. The 2014 transition to hybrid power units marked another profound technological shift. Most recently, the 2022 regulations---in the focus of the recent study \citet{LeePrinceZeager2026}---reintroduced ground-effect aerodynamics with the explicit goal of improving wheel-to-wheel racing and reducing turbulence.

Most of these regulatory interventions have empirically observable effects on competitive balance, especially in the first year when they become effective. However, the impact of the rule changes appears to be more moderate in recent years.

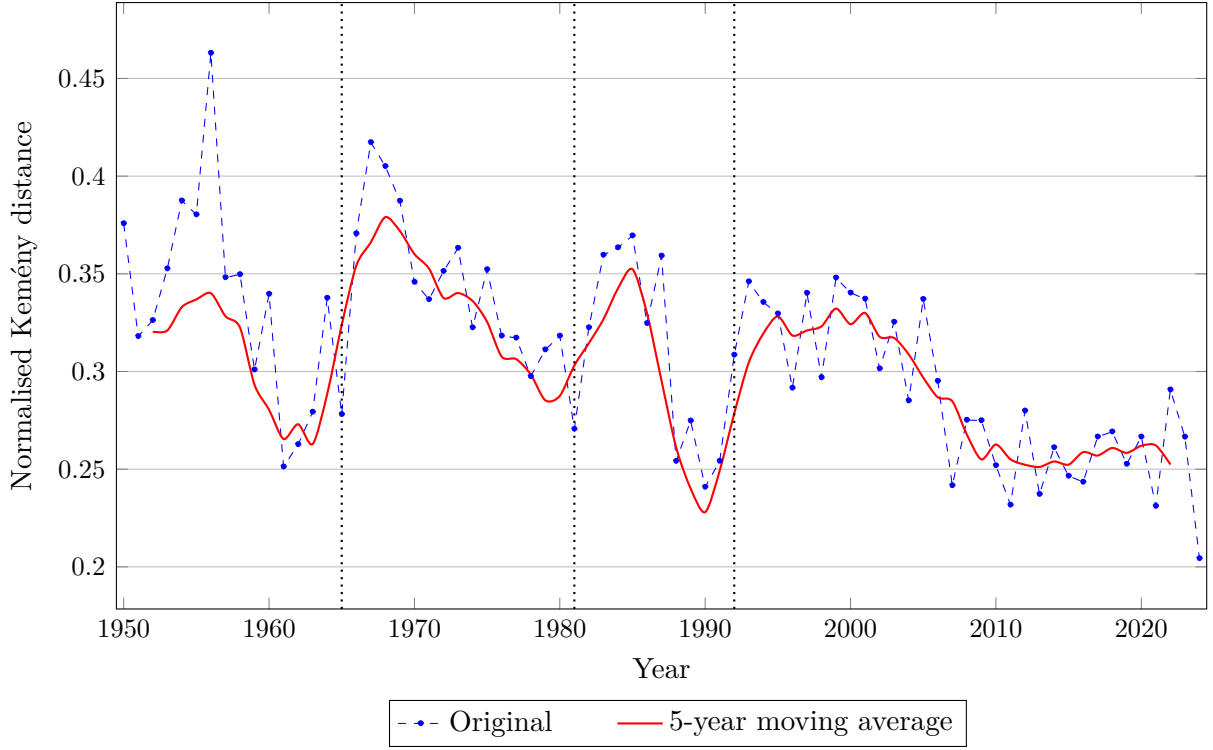
\begin{figure}[t!]
\centering
\begin{tikzpicture}
\begin{axis}[
width = \textwidth, 
height = 0.6\textwidth,
xlabel style = {align=center, font=\small},
xlabel = {Year},
ylabel style = {align=center, font=\small},
ylabel = {Normalised Kem\'eny distance},
xmin = 1949.5,
xmax = 2024.5,
ymajorgrids,
xtick = {1950,1960,1970,1980,1990,2000,2010,2020}, 
scaled x ticks = false,
xticklabel style = {/pgf/number format/1000 sep=},
legend style = {font=\small,at={(0.25,-0.15)},anchor=north west,legend columns=3},
legend entries = {Original$\qquad$, 5-year moving average},
]

\addplot[blue, dashed, thin, mark=*, mark size=1pt] coordinates {
(1950,0.3757494839290601)
(1951,0.318000962458012)
(1952,0.3261808614901992)
(1953,0.3526433330319852)
(1954,0.387412159087406)
(1955,0.3803016104965535)
(1956,0.4629685864244688)
(1957,0.3481103865091027)
(1958,0.3496728367387577)
(1959,0.3009120477734854)
(1960,0.3396461091405566)
(1961,0.2512532334108537)
(1962,0.2626777058665218)
(1963,0.2792792422470084)
(1964,0.337592239045261)
(1965,0.2780730688006338)
(1966,0.3705751058176239)
(1967,0.4172772714938551)
(1968,0.4050230771668717)
(1969,0.3872683130423068)
(1970,0.3457990625734499)
(1971,0.3368856775862364)
(1972,0.3513346338007395)
(1973,0.3631633110735665)
(1974,0.3224831801644054)
(1975,0.3522113142642164)
(1976,0.3182319261198571)
(1977,0.3172028447725246)
(1978,0.2974561589330143)
(1979,0.3111783269944189)
(1980,0.3182155738805985)
(1981,0.2705558890634901)
(1982,0.3225059093437152)
(1983,0.3596200206774919)
(1984,0.3634371184371184)
(1985,0.369522692484649)
(1986,0.324642094017094)
(1987,0.3591927537036232)
(1988,0.2541110493140526)
(1989,0.2747861728124886)
(1990,0.2408673882954435)
(1991,0.2541620845552297)
(1992,0.3084637089983308)
(1993,0.3460054347826087)
(1994,0.3354522792022792)
(1995,0.3295691520755459)
(1996,0.2915854978354978)
(1997,0.34012577362961)
(1998,0.2969155844155844)
(1999,0.3479437229437229)
(2000,0.3402597402597402)
(2001,0.3371530430353959)
(2002,0.3014782946671492)
(2003,0.325328947368421)
(2004,0.2850877192982456)
(2005,0.336995998768852)
(2006,0.2950937950937951)
(2007,0.2416603004838299)
(2008,0.2751601225285436)
(2009,0.274922600619195)
(2010,0.251906941266209)
(2011,0.2316933638443936)
(2012,0.279891304347826)
(2013,0.2371838687628161)
(2014,0.2610964976941838)
(2015,0.2465319645863886)
(2016,0.243454957740672)
(2017,0.266578947368421)
(2018,0.269172932330827)
(2019,0.2526315789473684)
(2020,0.26656346749226)
(2021,0.2311004784688995)
(2022,0.2906698564593301)
(2023,0.2665071770334928)
(2024,0.204312865497076)
};

\addplot[red, thick, smooth] coordinates {
(1952,0.320419431538093)
(1953,0.320991851715304)
(1954,0.332949508259373)
(1955,0.336873430628032)
(1956,0.340057712249406)
(1957,0.32832280541766)
(1958,0.322471075443989)
(1959,0.2932060215682)
(1960,0.280423065205138)
(1961,0.265452208961148)
(1962,0.273001194205286)
(1963,0.262861338266892)
(1964,0.288495604190428)
(1965,0.323813509169969)
(1966,0.354198011033197)
(1967,0.366134266848992)
(1968,0.379038372457122)
(1969,0.371896020405386)
(1970,0.360123586257956)
(1971,0.352721353835641)
(1972,0.337655252899345)
(1973,0.340180826804822)
(1974,0.336124708845679)
(1975,0.325400886916091)
(1976,0.307636796660317)
(1977,0.306337607972941)
(1978,0.298505446678582)
(1979,0.28527829009603)
(1980,0.287504940592882)
(1981,0.303078046263522)
(1982,0.314521438354615)
(1983,0.326727813080212)
(1984,0.34250695152509)
(1985,0.352366024661477)
(1986,0.329738858282215)
(1987,0.295137080246227)
(1988,0.261217195312471)
(1989,0.240107087704498)
(1990,0.227992639780271)
(1991,0.249270578509417)
(1992,0.278514196187245)
(1993,0.304605998457241)
(1994,0.319396453495561)
(1995,0.328226667419442)
(1996,0.318447158884498)
(1997,0.321001946689286)
(1998,0.323089134863431)
(1999,0.332202643903411)
(2000,0.324218500797448)
(2001,0.329966962861699)
(2002,0.317774550011392)
(2003,0.317172731175908)
(2004,0.308760881587588)
(2005,0.296899141676313)
(2006,0.286799587234653)
(2007,0.284766563498843)
(2008,0.267748751998314)
(2009,0.25503052692311)
(2010,0.262640495811851)
(2011,0.255045245058706)
(2012,0.252280024473703)
(2013,0.251149627475689)
(2014,0.253952371206841)
(2015,0.252326131695018)
(2016,0.25872394440862)
(2017,0.256980835345974)
(2018,0.260856779075266)
(2019,0.258356372213957)
(2020,0.26198316647233)
(2021,0.262167718762146)
(2022,0.252378662788879)
};

\draw [black,thick,dotted] (1965,\pgfkeysvalueof{/pgfplots/ymin}) -- (1965,\pgfkeysvalueof{/pgfplots/ymax});
\draw [black,thick,dotted] (1981,\pgfkeysvalueof{/pgfplots/ymin}) -- (1981,\pgfkeysvalueof{/pgfplots/ymax});
\draw [black,thick,dotted] (1992,\pgfkeysvalueof{/pgfplots/ymin}) -- (1992,\pgfkeysvalueof{/pgfplots/ymax});

\end{axis}
\end{tikzpicture}

\caption{Normalised Kem\'eny distance of start--finish rankings with structural breaks}
\label{Fig3}
\end{figure}


As described in Section~\ref{Sec33}, structural breaks are identified using the method of \citet{BaiPerron2003}.
Figure~\ref{Fig3} presents the 5-year moving average of competitive balance, as well as the three breakpoints that are detected. 
The moving average highlights long-term structural changes, smoothing out short-term fluctuations.
The first breakpoint coincides with the transition from the 1.5-litre engine to the subsequent 3.0-litre era: the 1965 season was the final year of the 1.5-litre regulation, the following season allowed 3.0-litre naturally aspirated engines and 1.5-litre supercharged engines.
The second break in the early 1980s occurred in a period marked by both technical and institutional changes after more than a decade of worsening competitive balance. In 1981, restrictions on sliding skirts were introduced to curb the extreme performance of ground-effect cars, while the first Concorde Agreement reshaped the commercial and organisational structure of Formula One.
In contrast, Formula One did not undergo a single technical reset, comparable to the 1966 engine regulation or the 1981 ground-effect restrictions, around the third breakpoint in 1992. Rather, it seems that competitive balance has become less volatile due to the continuously decreasing rate of technical failures.

\citet{PeetersWesselbaum2023} found three breakpoints visually in 2005, 2011, and 2019. Our results with longer data are different, and we do not detect any breakpoint in the 1993--2019 period analysed by them.

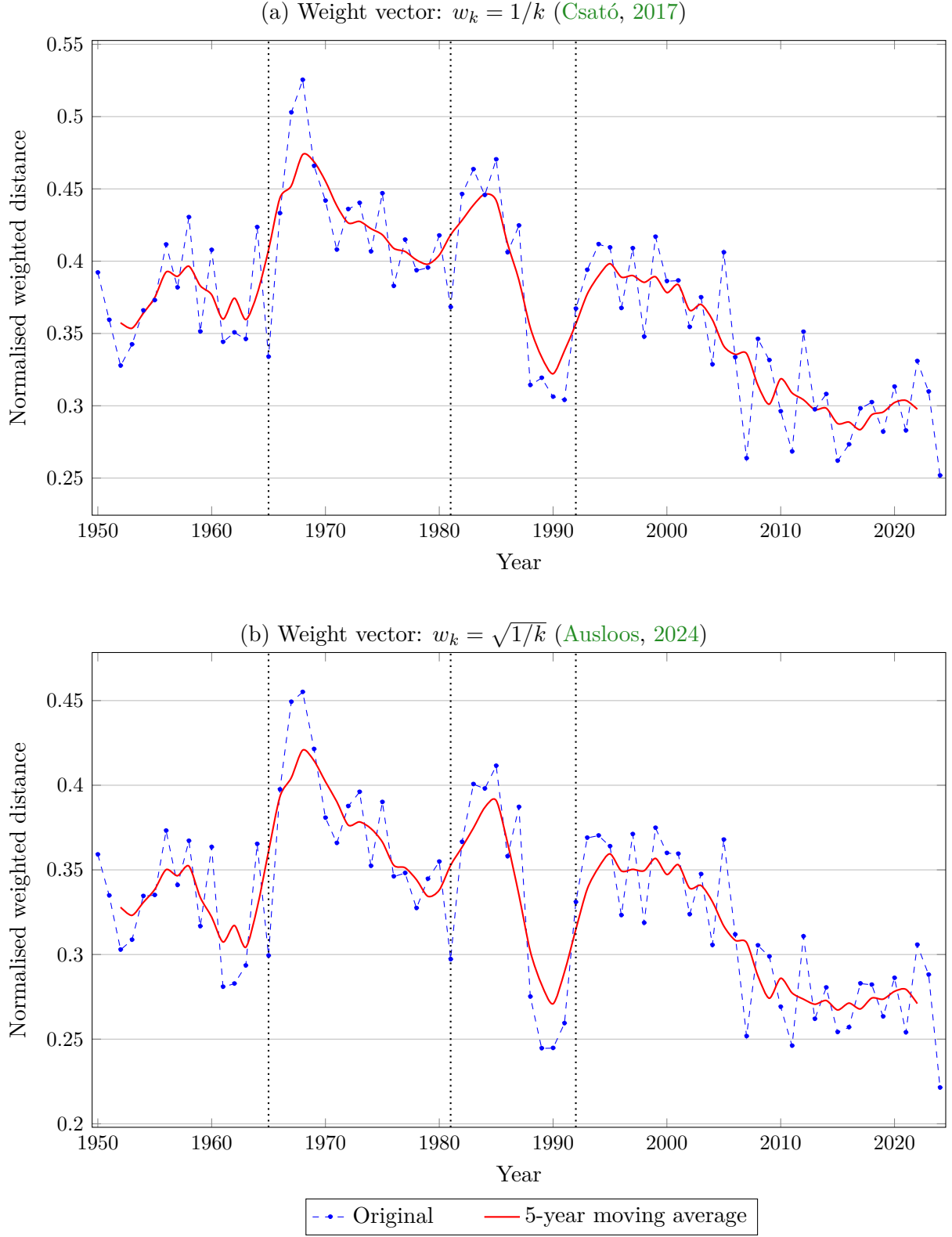
\begin{figure}[t!]
\centering

\begin{subfigure}{\textwidth}
\caption{Weight vector: $w_k = 1/k$ \citep{Csato2017a}}
\label{Fig4a}

\begin{tikzpicture}
\begin{axis}[
width = \textwidth, 
height = 0.6\textwidth,
xlabel style = {align=center, font=\small},
xlabel = {Year},
ylabel style = {align=center, font=\small},
ylabel = {Normalised weighted distance},
xmin = 1949.5,
xmax = 2024.5,
ymajorgrids,
xtick = {1950,1960,1970,1980,1990,2000,2010,2020}, 
scaled x ticks = false,
xticklabel style = {/pgf/number format/1000 sep=},
]

\addplot[blue, dashed, thin, mark=*, mark size=1pt] coordinates {
(1950,0.392047484773701)
(1951,0.35934258658073)
(1952,0.327650086118243)
(1953,0.342356055614984)
(1954,0.365841916385621)
(1955,0.372939439099602)
(1956,0.411443549603954)
(1957,0.381754739350068)
(1958,0.430351530068883)
(1959,0.351264838133085)
(1960,0.407696149198077)
(1961,0.34410862392228)
(1962,0.350634719647818)
(1963,0.34606488185752)
(1964,0.423445566872684)
(1965,0.333809623086361)
(1966,0.433069582761568)
(1967,0.502780498820632)
(1968,0.525291392117643)
(1969,0.465781980473744)
(1970,0.441777620110936)
(1971,0.407882446298514)
(1972,0.43589591244278)
(1973,0.440202945975931)
(1974,0.406632710672415)
(1975,0.446858198780162)
(1976,0.382761426751792)
(1977,0.414765453048297)
(1978,0.393583877469657)
(1979,0.395451946004263)
(1980,0.417658327170299)
(1981,0.368176330606151)
(1982,0.446304237225173)
(1983,0.463489348465983)
(1984,0.44566671569459)
(1985,0.470313208333066)
(1986,0.406019694396974)
(1987,0.424569947569406)
(1988,0.314228380965209)
(1989,0.319232305801358)
(1990,0.306172493493562)
(1991,0.303923823712997)
(1992,0.367070316223523)
(1993,0.393909676539973)
(1994,0.411660404987907)
(1995,0.409395236678688)
(1996,0.367473723156776)
(1997,0.408853079256158)
(1998,0.347689000759665)
(1999,0.416803363922655)
(2000,0.386105769007562)
(2001,0.386530967579826)
(2002,0.354426194169658)
(2003,0.374936937530631)
(2004,0.328511117842903)
(2005,0.405997799505231)
(2006,0.333457723172512)
(2007,0.263601111034573)
(2008,0.346136296413854)
(2009,0.331508313548062)
(2010,0.296066226760976)
(2011,0.268280790271274)
(2012,0.351078472932952)
(2013,0.297404763270763)
(2014,0.30800273497165)
(2015,0.261910319549702)
(2016,0.273200541099806)
(2017,0.298086706612146)
(2018,0.302389666378877)
(2019,0.281970866001278)
(2020,0.313150287078796)
(2021,0.282796284629613)
(2022,0.330813083999512)
(2023,0.30973406419212)
(2024,0.251682363937534)
};

\addplot[red, thick, smooth] coordinates {
(1952,0.357447625894656)
(1953,0.353626016759836)
(1954,0.364046209364481)
(1955,0.374867140010846)
(1956,0.392466234901626)
(1957,0.389550819251118)
(1958,0.396502161270813)
(1959,0.383035176134479)
(1960,0.376811172194029)
(1961,0.359953842551756)
(1962,0.374389988299676)
(1963,0.359612683077333)
(1964,0.37740487484519)
(1965,0.407834030679753)
(1966,0.443679332731778)
(1967,0.45214661545199)
(1968,0.473740214856905)
(1969,0.468702787564294)
(1970,0.455325870288723)
(1971,0.438308181060381)
(1972,0.426478327100115)
(1973,0.42749444283396)
(1974,0.422470238924616)
(1975,0.418244147045719)
(1976,0.408920333344465)
(1977,0.406684180410834)
(1978,0.400844206088862)
(1979,0.397927186859733)
(1980,0.404234943695109)
(1981,0.418216037894374)
(1982,0.428258991832439)
(1983,0.438789968064993)
(1984,0.446358640823157)
(1985,0.442011782892004)
(1986,0.412159589391849)
(1987,0.386872707413203)
(1988,0.354044564445302)
(1989,0.333625390308506)
(1990,0.32212546403933)
(1991,0.338061723154283)
(1992,0.356547342991592)
(1993,0.377191891628618)
(1994,0.389901871517373)
(1995,0.3982584241239)
(1996,0.389014288967839)
(1997,0.390042880754788)
(1998,0.385384987220563)
(1999,0.389196436105173)
(2000,0.378311059087873)
(2001,0.383760646442066)
(2002,0.366102197226116)
(2003,0.37008060332565)
(2004,0.359465954444187)
(2005,0.34130093781717)
(2006,0.335540809593815)
(2007,0.336140248734846)
(2008,0.314153934185995)
(2009,0.301118547605748)
(2010,0.318614019985424)
(2011,0.308867713356805)
(2012,0.304166597641523)
(2013,0.297335416199268)
(2014,0.298319366364975)
(2015,0.287721013100813)
(2016,0.288717993722436)
(2017,0.283511619928362)
(2018,0.293759613434181)
(2019,0.295678762140142)
(2020,0.302224037617615)
(2021,0.303692917180264)
(2022,0.297635216767515)
};

\draw [black,thick,dotted] (1965,\pgfkeysvalueof{/pgfplots/ymin}) -- (1965,\pgfkeysvalueof{/pgfplots/ymax});
\draw [black,thick,dotted] (1981,\pgfkeysvalueof{/pgfplots/ymin}) -- (1981,\pgfkeysvalueof{/pgfplots/ymax});
\draw [black,thick,dotted] (1992,\pgfkeysvalueof{/pgfplots/ymin}) -- (1992,\pgfkeysvalueof{/pgfplots/ymax});

\end{axis}
\end{tikzpicture}
\end{subfigure}

\vspace{0.5cm}
\begin{subfigure}{\textwidth}
\caption{Weight vector: $w_k = \sqrt{1/k}$ \citep{Ausloos2024a}}
\label{Fig4b}

\begin{tikzpicture}
\begin{axis}[
width = \textwidth, 
height = 0.6\textwidth,
xlabel style = {align=center, font=\small},
xlabel = {Year},
ylabel style = {align=center, font=\small},
ylabel = {Normalised weighted distance},
xmin = 1949.5,
xmax = 2024.5,
ymajorgrids,
xtick = {1950,1960,1970,1980,1990,2000,2010,2020}, 
scaled x ticks = false,
xticklabel style = {/pgf/number format/1000 sep=},
legend style = {font=\small,at={(0.25,-0.15)},anchor=north west,legend columns=3},
legend entries = {Original$\qquad$, 5-year moving average},
]

\addplot[blue, dashed, thin, mark=*, mark size=1pt] coordinates {
(1950,0.359026209046613)
(1951,0.334780935605552)
(1952,0.302777059785179)
(1953,0.30862628446185)
(1954,0.33447139418776)
(1955,0.335052242223678)
(1956,0.373081516782127)
(1957,0.340946338484593)
(1958,0.367061125334154)
(1959,0.316615748605751)
(1960,0.363429681936016)
(1961,0.2808597858144)
(1962,0.282740651432845)
(1963,0.293411700291333)
(1964,0.365246614073598)
(1965,0.299242593139757)
(1966,0.397423785675981)
(1967,0.449204189585092)
(1968,0.455017781322809)
(1969,0.421291136299658)
(1970,0.380729128761417)
(1971,0.365799285555488)
(1972,0.387577341404768)
(1973,0.395990609485041)
(1974,0.35226838523562)
(1975,0.39000765725945)
(1976,0.346056790334089)
(1977,0.348047285582402)
(1978,0.327354117909039)
(1979,0.344639164372693)
(1980,0.354795688333443)
(1981,0.29712008690337)
(1982,0.366452204597107)
(1983,0.400590248185069)
(1984,0.397947258988632)
(1985,0.411391746630481)
(1986,0.357926765072449)
(1987,0.387100635806163)
(1988,0.275037862598645)
(1989,0.244531181874498)
(1990,0.244702031038653)
(1991,0.259316364485245)
(1992,0.330895018765082)
(1993,0.36889730298746)
(1994,0.370211740070539)
(1995,0.363888846334026)
(1996,0.323163737464267)
(1997,0.371036021971778)
(1998,0.318629234348206)
(1999,0.374769971029191)
(2000,0.359897569551154)
(2001,0.359443918252596)
(2002,0.323678045946628)
(2003,0.347420944388756)
(2004,0.305524697739735)
(2005,0.367787507311162)
(2006,0.311735714462809)
(2007,0.25167329561579)
(2008,0.305318355979141)
(2009,0.298780135212542)
(2010,0.269022816636581)
(2011,0.246049092100206)
(2012,0.310664823112854)
(2013,0.261911430473316)
(2014,0.280430630233803)
(2015,0.254096078132341)
(2016,0.256952300917775)
(2017,0.282823876797396)
(2018,0.282108706748956)
(2019,0.263311178497726)
(2020,0.286101517221968)
(2021,0.253903282262436)
(2022,0.305699570485391)
(2023,0.288012436578703)
(2024,0.221305227696021)
};

\addplot[red, thick, smooth] coordinates {
(1952,0.327936376617391)
(1953,0.323141583252804)
(1954,0.330801699488119)
(1955,0.338435555228002)
(1956,0.350122523402462)
(1957,0.346551394286061)
(1958,0.352226882228528)
(1959,0.333782536034983)
(1960,0.322141398624633)
(1961,0.307411513616069)
(1962,0.317137686709638)
(1963,0.304300268950387)
(1964,0.327613068922703)
(1965,0.360905776553152)
(1966,0.393226992759447)
(1967,0.404435897204659)
(1968,0.420733204328991)
(1969,0.414408304304893)
(1970,0.402082934668828)
(1971,0.390277500301274)
(1972,0.376472950088467)
(1973,0.378328655788073)
(1974,0.374380156743794)
(1975,0.36647414557932)
(1976,0.35274684726412)
(1977,0.351221003091535)
(1978,0.344178609306333)
(1979,0.334391268620189)
(1980,0.33807225242313)
(1981,0.352719478478336)
(1982,0.363381097401524)
(1983,0.374700309060932)
(1984,0.386861644694748)
(1985,0.390991330936559)
(1986,0.365880853819274)
(1987,0.335197638396447)
(1988,0.301859695278082)
(1989,0.282137615160641)
(1990,0.270896491752425)
(1991,0.289668379830188)
(1992,0.314804491469396)
(1993,0.33864185452847)
(1994,0.351411329124275)
(1995,0.359439529765614)
(1996,0.349385916037763)
(1997,0.350297562229494)
(1998,0.349499306872919)
(1999,0.356755343030585)
(2000,0.347283747825555)
(2001,0.353042089833665)
(2002,0.339193035175774)
(2003,0.340771022727775)
(2004,0.331229381969818)
(2005,0.31682843190365)
(2006,0.308407914221727)
(2007,0.307059001716289)
(2008,0.287306063581373)
(2009,0.274168739108852)
(2010,0.285967044608265)
(2011,0.2772856595071)
(2012,0.273615758511352)
(2013,0.270630410810504)
(2014,0.272811052574018)
(2015,0.267242863310926)
(2016,0.271282318566054)
(2017,0.267858428218839)
(2018,0.274259516036764)
(2019,0.273649712305696)
(2020,0.278224851043295)
(2021,0.279405597009245)
(2022,0.271004406848904)
};

\draw [black,thick,dotted] (1965,\pgfkeysvalueof{/pgfplots/ymin}) -- (1965,\pgfkeysvalueof{/pgfplots/ymax});
\draw [black,thick,dotted] (1981,\pgfkeysvalueof{/pgfplots/ymin}) -- (1981,\pgfkeysvalueof{/pgfplots/ymax});
\draw [black,thick,dotted] (1992,\pgfkeysvalueof{/pgfplots/ymin}) -- (1992,\pgfkeysvalueof{/pgfplots/ymax});

\end{axis}
\end{tikzpicture}
\end{subfigure}

\caption{Normalised weighted distance of start--finish rankings with structural breaks}
\label{Fig4}
\end{figure}


Figure~\ref{Fig4} repeats this investigation by using weighted distances. The main disagreement is that in the early years, the new measures are less volatile since most of the swaps were at the bottom of the ranking. The patterns of the two weighting schemes, proposed by \citet{Csato2017a} and \citet{Ausloos2024a}, are almost identical.
However, the absolute value of the indicator is consistently higher for the top-heavy weighting $w_k = 1/k$. Consequently, the competition turns out to be more balanced if one focuses on the top positions.

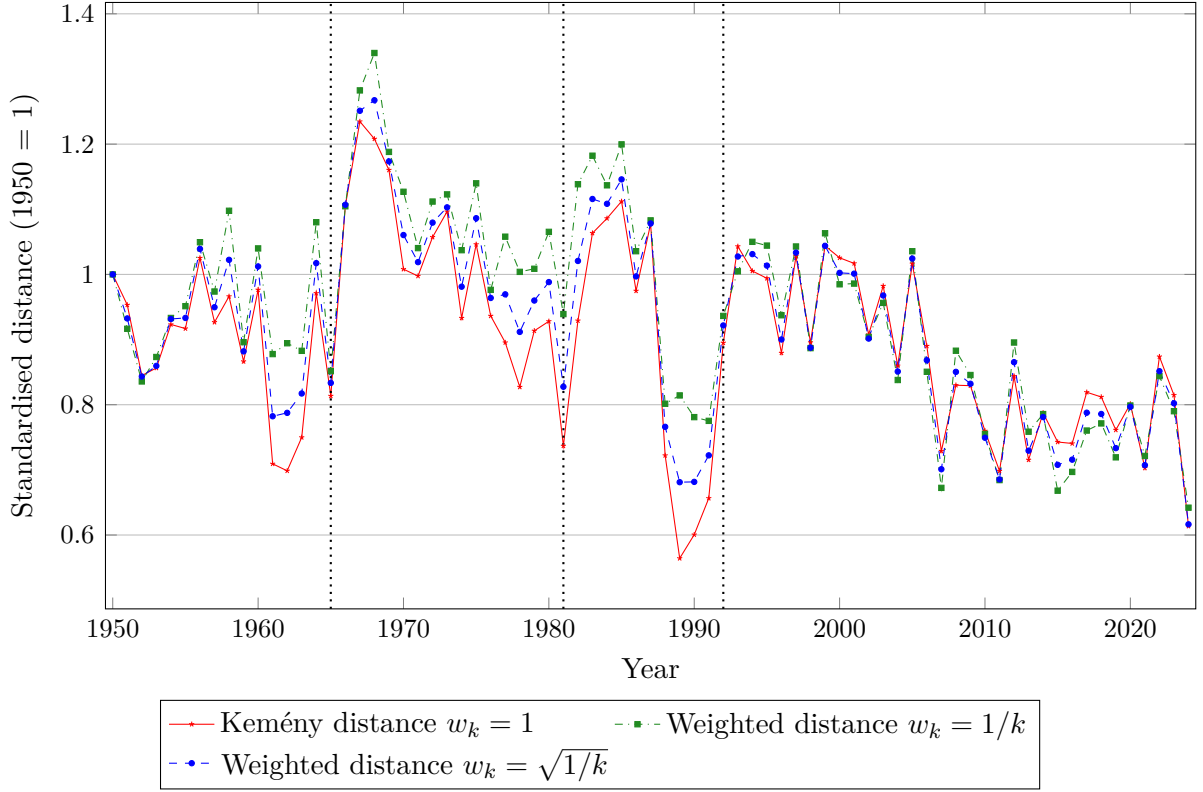
\begin{figure}[t!]
\centering
\begin{tikzpicture}
\begin{axis}[
width = \textwidth, 
height = 0.6\textwidth,
xlabel style = {align=center, font=\small},
xlabel = {Year},
ylabel style = {align=center, font=\small},
ylabel = {Standardised distance (1950 = 1)},
xmin = 1949.5,
xmax = 2024.5,
ymajorgrids,
xtick = {1950,1960,1970,1980,1990,2000,2010,2020}, 
scaled x ticks = false,
xticklabel style = {/pgf/number format/1000 sep=},
legend style = {font=\small,at={(0.05,-0.15)},anchor=north west,legend columns=2},
legend entries = {Kem\'eny distance $w_k = 1\quad \quad \;\;$, Weighted distance $w_k = 1/k$, Weighted distance $w_k = \sqrt{1/k}$},
]

\addplot[color=red, mark=star, mark size=1pt, mark options={solid}] coordinates {
(1950,1)
(1951,0.953192375362557)
(1952,0.844757581378279)
(1953,0.856382867442895)
(1954,0.923231085527933)
(1955,0.916922371180813)
(1956,1.02517149113246)
(1957,0.926624715584666)
(1958,0.966271236807772)
(1959,0.866410042142949)
(1960,0.976420120640172)
(1961,0.709278591366548)
(1962,0.698552111793174)
(1963,0.749841780609869)
(1964,0.971117111278136)
(1965,0.813445672766528)
(1966,1.10519323008646)
(1967,1.23473203395766)
(1968,1.20807405347407)
(1969,1.16026927137738)
(1970,1.00805401845243)
(1971,0.997478572495713)
(1972,1.05719055966251)
(1973,1.09644010922134)
(1974,0.933055617972207)
(1975,1.04614249777662)
(1976,0.936307776943752)
(1977,0.895463330474883)
(1978,0.827193630996207)
(1979,0.913462399436798)
(1980,0.928024745262255)
(1981,0.736827461234225)
(1982,0.929043710454107)
(1983,1.06339761108278)
(1984,1.08604155149208)
(1985,1.11211052883444)
(1986,0.97479451779427)
(1987,1.07772956062722)
(1988,0.722154633820294)
(1989,0.564208036960684)
(1990,0.600560724975447)
(1991,0.65643048361814)
(1992,0.895030136038923)
(1993,1.04304974902119)
(1994,1.00523450024132)
(1995,0.994054286631398)
(1996,0.879487088574616)
(1997,1.02819963851854)
(1998,0.89556382203005)
(1999,1.04376355784257)
(2000,1.02553141879696)
(2001,1.01692899826781)
(2002,0.907789830295021)
(2003,0.98225705955729)
(2004,0.859888334965977)
(2005,1.01645531762942)
(2006,0.890068897904844)
(2007,0.728901525193465)
(2008,0.829944482324671)
(2009,0.829228063112893)
(2010,0.759807685946527)
(2011,0.698263838561)
(2012,0.843668367711101)
(2013,0.715399605753913)
(2014,0.787525443819007)
(2015,0.742760034878542)
(2016,0.740532348868088)
(2017,0.819142856939898)
(2018,0.81188577736568)
(2019,0.761237409503946)
(2020,0.801213601521239)
(2021,0.702823416719044)
(2022,0.873838970889902)
(2023,0.814669032316018)
(2024,0.613607496464181)
};
\addplot[color=ForestGreen, dashdotted, mark=square*, mark size=1pt, mark options={solid}] coordinates {
(1950,1)
(1951,0.916579242405167)
(1952,0.835740819271855)
(1953,0.873251503737105)
(1954,0.933157157217304)
(1955,0.951260889519216)
(1956,1.04947376423407)
(1957,0.973746176615381)
(1958,1.09770256609934)
(1959,0.895975237121706)
(1960,1.03991522719093)
(1961,0.877721799748078)
(1962,0.894367986699909)
(1963,0.882711649220953)
(1964,1.08008744684871)
(1965,0.851452020611849)
(1966,1.10463553416634)
(1967,1.28244796446239)
(1968,1.33986675726501)
(1969,1.18807542087052)
(1970,1.1268472245548)
(1971,1.04039041733415)
(1972,1.11184468558544)
(1973,1.12283068524219)
(1974,1.03720270238982)
(1975,1.13980631463074)
(1976,0.976313945676073)
(1977,1.05794698131455)
(1978,1.00391889440853)
(1979,1.00868379816931)
(1980,1.06532586839929)
(1981,0.939111574248897)
(1982,1.13839331856137)
(1983,1.18222757820655)
(1984,1.13676718510728)
(1985,1.19963327555728)
(1986,1.03563907477008)
(1987,1.08295541754203)
(1988,0.801505922545553)
(1989,0.814269490813406)
(1990,0.780957678303413)
(1991,0.775221970594784)
(1992,0.936290450722837)
(1993,1.00474991382064)
(1994,1.05002689974029)
(1995,1.04424910904607)
(1996,0.937319425397897)
(1997,1.04286622191227)
(1998,0.8868543078662)
(1999,1.06314510387241)
(2000,0.984844397689304)
(2001,0.985928956547039)
(2002,0.904038944094339)
(2003,0.956355931596024)
(2004,0.837937062732407)
(2005,1.03558322721948)
(2006,0.850554425479841)
(2007,0.672370366530294)
(2008,0.882893807145969)
(2009,0.845582044071565)
(2010,0.755179508247253)
(2011,0.684306877841933)
(2012,0.895499873275817)
(2013,0.758593728620481)
(2014,0.785626096158827)
(2015,0.668057645366308)
(2016,0.696855742506558)
(2017,0.760333169295063)
(2018,0.771308777949241)
(2019,0.719226310466035)
(2020,0.798756016148283)
(2021,0.72133171519478)
(2022,0.843808714116518)
(2023,0.790042217388299)
(2024,0.641969082094254)
};
\addplot[blue, dashed, mark=*, mark size=1pt, mark options={solid}] coordinates {
(1950,1)
(1951,0.932469349506701)
(1952,0.843328570883996)
(1953,0.859620486430227)
(1954,0.931607180088446)
(1955,0.93322502307952)
(1956,1.03914841697167)
(1957,0.949641919986759)
(1958,1.02237974856732)
(1959,0.8818736365975)
(1960,1.01226504577785)
(1961,0.782282125196981)
(1962,0.787520922730564)
(1963,0.817243122919861)
(1964,1.01732576862147)
(1965,0.833483978605322)
(1966,1.10694923006132)
(1967,1.25117380922676)
(1968,1.26736647592135)
(1969,1.17342724760509)
(1970,1.0604494022106)
(1971,1.01886513111915)
(1972,1.07952381090498)
(1973,1.102957387252)
(1974,0.981177352402939)
(1975,1.08629299876215)
(1976,0.963876122729413)
(1977,0.969420272984066)
(1978,0.911783345227975)
(1979,0.959927592160682)
(1980,0.98821667999001)
(1981,0.827572136564533)
(1982,1.02068371434557)
(1983,1.11576881601159)
(1984,1.10840726654851)
(1985,1.1458543589977)
(1986,0.996937705531071)
(1987,1.07819603709184)
(1988,0.76606625273682)
(1989,0.681095629547062)
(1990,0.681571497770189)
(1991,0.722276975750195)
(1992,0.921645858790552)
(1993,1.02749407617639)
(1994,1.03115519352648)
(1995,1.01354396187489)
(1996,0.900111828388301)
(1997,1.03345107577816)
(1998,0.887481822550837)
(1999,1.04385128880809)
(2000,1.00242701084931)
(2001,1.00116345045419)
(2002,0.901544337963929)
(2003,0.967675717355915)
(2004,0.850981599786405)
(2005,1.02440294898753)
(2006,0.868281219052551)
(2007,0.700988644489502)
(2008,0.850406873609333)
(2009,0.832195889001939)
(2010,0.749312473178395)
(2011,0.685323483078254)
(2012,0.865298452549797)
(2013,0.729505044127048)
(2014,0.781086793018485)
(2015,0.707736849649746)
(2016,0.715692321182085)
(2017,0.787752731335212)
(2018,0.785760759633928)
(2019,0.733403779063773)
(2020,0.796881982465027)
(2021,0.707199853004245)
(2022,0.851468674939276)
(2023,0.802204489035813)
(2024,0.6164041012039)
};

\draw [black,thick,dotted] (1965,\pgfkeysvalueof{/pgfplots/ymin}) -- (1965,\pgfkeysvalueof{/pgfplots/ymax});
\draw [black,thick,dotted] (1981,\pgfkeysvalueof{/pgfplots/ymin}) -- (1981,\pgfkeysvalueof{/pgfplots/ymax});
\draw [black,thick,dotted] (1992,\pgfkeysvalueof{/pgfplots/ymin}) -- (1992,\pgfkeysvalueof{/pgfplots/ymax});

\end{axis}
\end{tikzpicture}

\caption{Evolution of competitive balance based on start--finish rankings}
\label{Fig5}
\end{figure}


Figure~\ref{Fig5} aims to explore the evolution of competitive balance based on our three indicators, standardised to their level in 1950. Their movements are highly consistent, especially in the last three decades, although the Kem\'eny distance suggests a more unfavourable picture roughly until 1990. Crucially, competitiveness within races is worse in the last two decades---after the dominance of Michael Schumacher---than ever before. This key result has not been observed by \citet{PeetersWesselbaum2023} due to their restricted sample.

\begin{table}[t!]
  \centering
  \caption{Sensitivity analysis of the Bai--Perron test, start--finish positions}
  \label{Table1}
\centerline{
\begin{threeparttable}
    \rowcolors{1}{gray!20}{}
    \begin{tabularx}{1\textwidth}{lCCc} \toprule \hiderowcolors
    \multirow{2}[0]{*}{Min} & \multicolumn{3}{c}{Weight vector} \\
          & $w_k = 1$ (Kem\'eny) & $w_k = 1/k$ \citep{Csato2017a} & $w_k = \sqrt{1/k}$ \citep{Ausloos2024a} \\ \bottomrule \showrowcolors
    5\%   & 1965, 1981, 1987, 1990 & 1965, 1981, 1991 & 1965, 1982, 1991 \\
    10\%  & 1965, 1982, 1991 & 1965, 1981, 1991 & 1965, 1982, 1991 \\
    15\%  & 1965, 1981, 1992 & 1965, 1981, 1992 & 1965, 1982, 1992 \\
    20\%  & 1965, 1991 & 1967  & 1965, 1991 \\ \bottomrule
    \end{tabularx}
\begin{tablenotes} \footnotesize
\item
Min is the trimming parameter, the minimum distance required between two consecutive breakpoints as a proportion of the total sample size.
\item
The cells show the years of structural breaks.
\end{tablenotes}
\end{threeparttable}
}
\end{table}

Table~\ref{Table1} investigates the sensitivity of the structural breaks to the minimum distance of two consecutive breakpoints in the Bai--Perron test. Clearly, the existence of three structural breaks is difficult to deny.

\begin{figure}[t!]
\centering
\begin{tikzpicture}
\begin{axis}[
width = \textwidth, 
height = 0.6\textwidth,
xlabel style = {align=center, font=\small},
xlabel = {Year},
ylabel style = {align=center, font=\small},
ylabel = {Normalised Kem\'eny distance},
xmin = 1949.5,
xmax = 2024.5,
ymajorgrids,
xtick = {1950,1960,1970,1980,1990,2000,2010,2020}, 
scaled x ticks = false,
xticklabel style = {/pgf/number format/1000 sep=},
legend style = {font=\small,at={(0.25,-0.15)},anchor=north west,legend columns=3},
legend entries = {Start--start$\qquad$, Finish--finish},
]

\addplot[blue, dashed, mark=*, mark size=1pt, mark options={solid}] coordinates {
(1950,0.125487)
(1951,0.200722)
(1952,0.264523)
(1953,0.176800)
(1954,0.232480)
(1955,0.145066)
(1956,0.201405)
(1957,0.191389)
(1958,0.288788)
(1959,0.293349)
(1960,0.353654)
(1961,0.233146)
(1962,0.290596)
(1963,0.219834)
(1964,0.181056)
(1965,0.290252)
(1966,0.254037)
(1967,0.263118)
(1968,0.343500)
(1969,0.289511)
(1970,0.317243)
(1971,0.306426)
(1972,0.263021)
(1973,0.279324)
(1974,0.285038)
(1975,0.268956)
(1976,0.295133)
(1977,0.254401)
(1978,0.219438)
(1979,0.272965)
(1980,0.302060)
(1981,0.233685)
(1982,0.226575)
(1983,0.237634)
(1984,0.235125)
(1985,0.234105)
(1986,0.180118)
(1987,0.159272)
(1988,0.193869)
(1989,0.215281)
(1990,0.187036)
(1991,0.176134)
(1992,0.220955)
(1993,0.195737)
(1994,0.199258)
(1995,0.135541)
(1996,0.144584)
(1997,0.228245)
(1998,0.180344)
(1999,0.218190)
(2000,0.224563)
(2001,0.176393)
(2002,0.194767)
(2003,0.289855)
(2004,0.278905)
(2005,0.315479)
(2006,0.293778)
(2007,0.228476)
(2008,0.257773)
(2009,0.379198)
(2010,0.190798)
(2011,0.163161)
(2012,0.243062)
(2013,0.214212)
(2014,0.258327)
(2015,0.276294)
(2016,0.225335)
(2017,0.271222)
(2018,0.287669)
(2019,0.278045)
(2020,0.224989)
(2021,0.278682)
(2022,0.311521)
(2023,0.343125)
(2024,0.302974)
};

\addplot[color=ForestGreen, dashdotted, mark=square*, mark size=1pt, mark options={solid}] coordinates {
(1950,0.339253)
(1951,0.341220)
(1952,0.363353)
(1953,0.271066)
(1954,0.308859)
(1955,0.210340)
(1956,0.193305)
(1957,0.257505)
(1958,0.431332)
(1959,0.460742)
(1960,0.371391)
(1961,0.354385)
(1962,0.395098)
(1963,0.386242)
(1964,0.442395)
(1965,0.429157)
(1966,0.490411)
(1967,0.431948)
(1968,0.477560)
(1969,0.430748)
(1970,0.448432)
(1971,0.471185)
(1972,0.432440)
(1973,0.428849)
(1974,0.433524)
(1975,0.434971)
(1976,0.430060)
(1977,0.447145)
(1978,0.411598)
(1979,0.407605)
(1980,0.414115)
(1981,0.382205)
(1982,0.407160)
(1983,0.425223)
(1984,0.451197)
(1985,0.418260)
(1986,0.407197)
(1987,0.422431)
(1988,0.375188)
(1989,0.315594)
(1990,0.325660)
(1991,0.343718)
(1992,0.402588)
(1993,0.434443)
(1994,0.396885)
(1995,0.427008)
(1996,0.413216)
(1997,0.437045)
(1998,0.379427)
(1999,0.419310)
(2000,0.412322)
(2001,0.429573)
(2002,0.399572)
(2003,0.417199)
(2004,0.346116)
(2005,0.395605)
(2006,0.367152)
(2007,0.346917)
(2008,0.379351)
(2009,0.416550)
(2010,0.336303)
(2011,0.293435)
(2012,0.353315)
(2013,0.321088)
(2014,0.319831)
(2015,0.349042)
(2016,0.329945)
(2017,0.347032)
(2018,0.361253)
(2019,0.335539)
(2020,0.362107)
(2021,0.327440)
(2022,0.350038)
(2023,0.349497)
(2024,0.315935)
};

\draw [blue,thick,dotted] (1967,\pgfkeysvalueof{/pgfplots/ymin}) -- (1967,\pgfkeysvalueof{/pgfplots/ymax});
\draw [blue,thick,dotted] (1994,\pgfkeysvalueof{/pgfplots/ymin}) -- (1994,\pgfkeysvalueof{/pgfplots/ymax});
\draw [blue,thick,dotted] (2009,\pgfkeysvalueof{/pgfplots/ymin}) -- (2009,\pgfkeysvalueof{/pgfplots/ymax});

\draw [ForestGreen,thick,dotted] (1963,\pgfkeysvalueof{/pgfplots/ymin}) -- (1963,\pgfkeysvalueof{/pgfplots/ymax});

\end{axis}
\end{tikzpicture}
\caption{Normalised Kem\'eny distance of start--start \\ and finish--finish rankings with structural breaks}
\label{Fig6}
\end{figure}
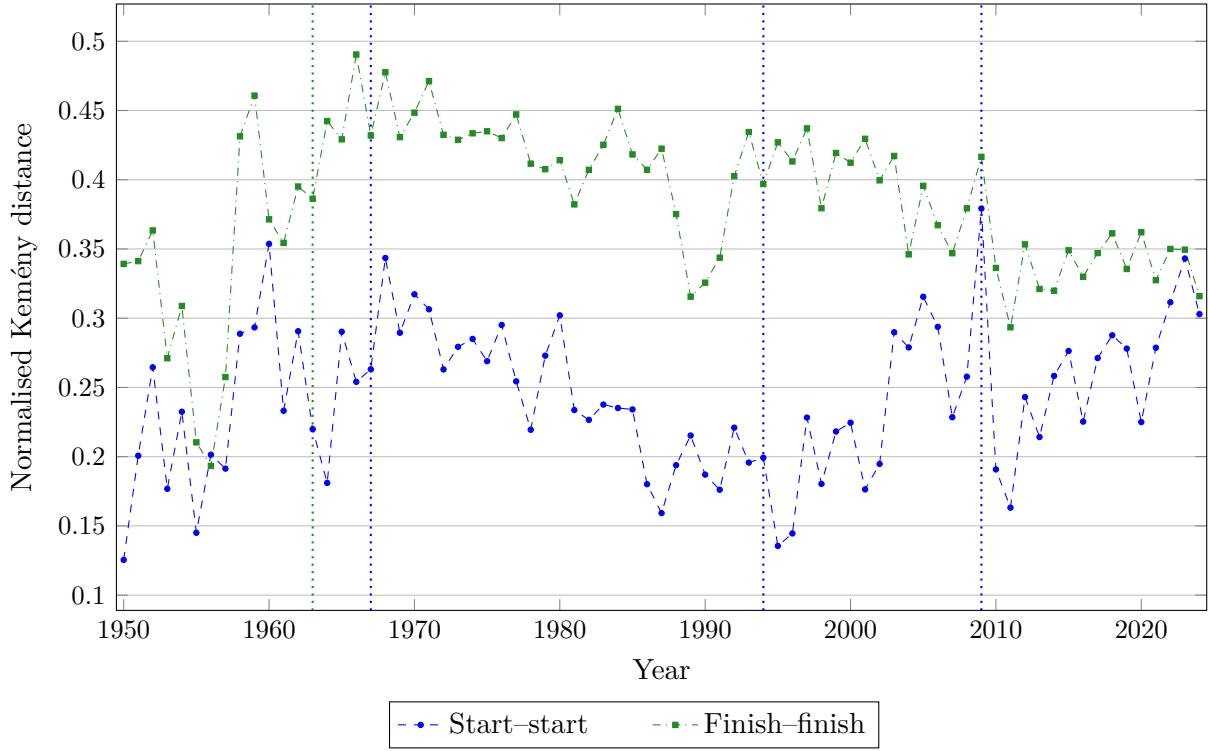


\citet{PeetersWesselbaum2023} also compute the average Kem\'eny distance between the start and finish positions over all races within the seasons, which they call mean competitiveness across seasons.
Figure~\ref{Fig6} substantially extends their dataset. The main conclusion remains the same: the finish positions always imply a higher competitive balance than the start positions. This is expected as retirement gives an additional randomness to finish rankings, and qualifying is more predictable than the race itself due to the minimal interaction between cars.

The time series based on finish positions varies less. We find only one structural break for the finish rankings (1963), and three for the start rankings (1967, 1994, 2009).
The Bai--Perron test reinforces the statement of \citet{PeetersWesselbaum2023} that competitiveness based on start positions increased between 1993 and 2005, as well as between 2011 and 2019. However, competitive balance certainly did not improve in recent decades based on finish positions.

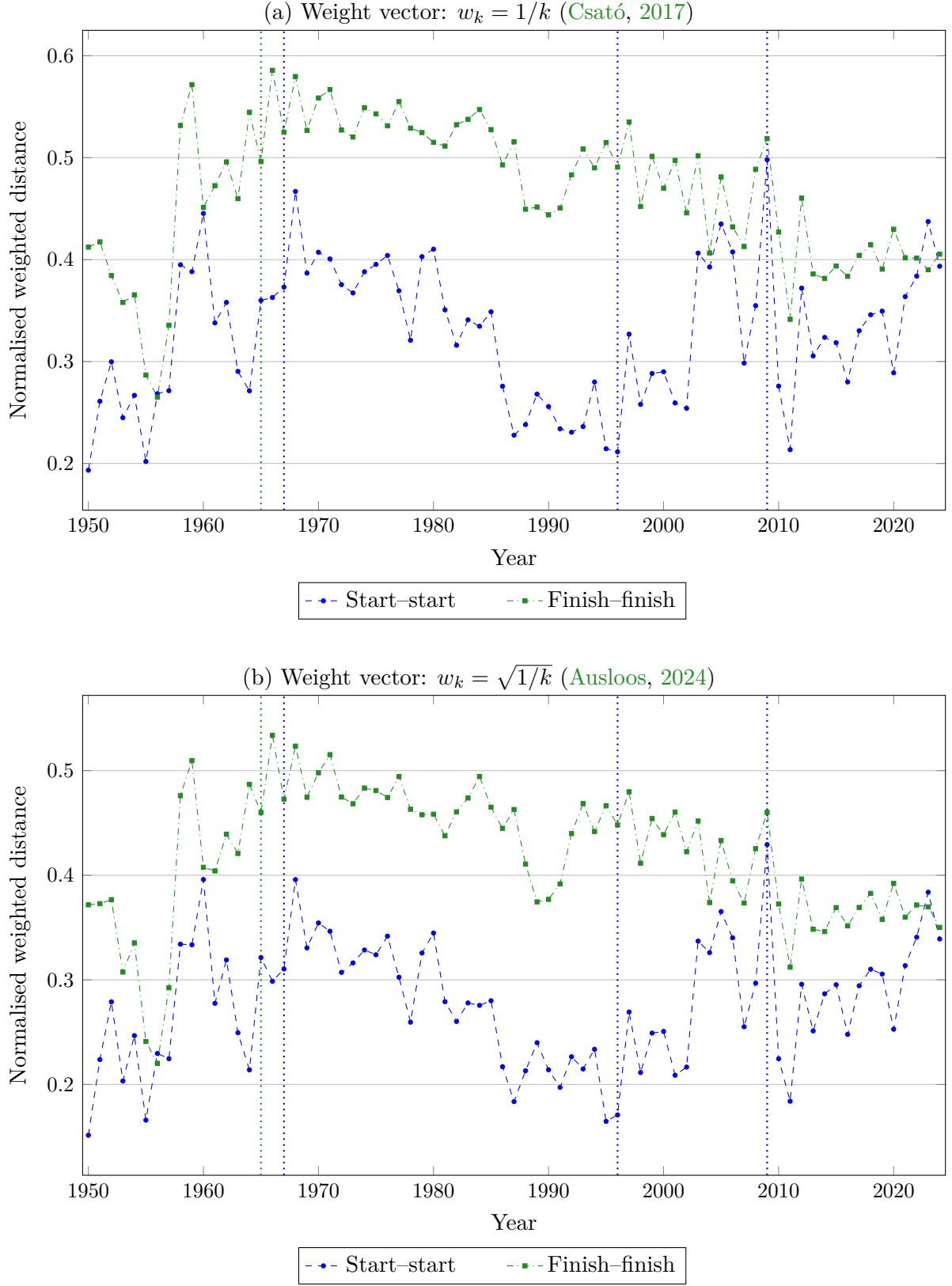
\begin{figure}[t!]
\centering

\begin{subfigure}{\textwidth}
\caption{Weight vector: $w_k = 1/k$ \citep{Csato2017a}}
\label{Fig6a}

\begin{tikzpicture}
\begin{axis}[
width = \textwidth, 
height = 0.6\textwidth,
xlabel style = {align=center, font=\small},
xlabel = {Year},
ylabel style = {align=center, font=\small},
ylabel = {Normalised weighted distance},
xmin = 1949.5,
xmax = 2024.5,
ymajorgrids,
xtick = {1950,1960,1970,1980,1990,2000,2010,2020}, 
scaled x ticks = false,
xticklabel style = {/pgf/number format/1000 sep=},
legend style = {font=\small,at={(0.25,-0.15)},anchor=north west,legend columns=3},
legend entries = {Start--start$\qquad$, Finish--finish},
]
\addplot[blue, dashed, mark=*, mark size=1pt, mark options={solid}] coordinates {
(1950,0.193377)
(1951,0.261028)
(1952,0.299741)
(1953,0.244885)
(1954,0.266807)
(1955,0.201958)
(1956,0.268463)
(1957,0.271423)
(1958,0.394986)
(1959,0.388133)
(1960,0.445334)
(1961,0.337913)
(1962,0.357991)
(1963,0.290332)
(1964,0.271313)
(1965,0.359930)
(1966,0.362833)
(1967,0.373001)
(1968,0.466835)
(1969,0.386715)
(1970,0.407242)
(1971,0.400679)
(1972,0.375467)
(1973,0.367299)
(1974,0.388072)
(1975,0.395350)
(1976,0.404093)
(1977,0.369400)
(1978,0.320853)
(1979,0.402896)
(1980,0.410286)
(1981,0.350697)
(1982,0.315972)
(1983,0.340872)
(1984,0.334607)
(1985,0.348788)
(1986,0.275762)
(1987,0.227730)
(1988,0.238199)
(1989,0.268075)
(1990,0.255846)
(1991,0.234025)
(1992,0.230679)
(1993,0.236162)
(1994,0.279941)
(1995,0.214438)
(1996,0.211406)
(1997,0.326867)
(1998,0.258017)
(1999,0.288319)
(2000,0.289997)
(2001,0.259343)
(2002,0.254128)
(2003,0.406408)
(2004,0.392737)
(2005,0.434964)
(2006,0.407536)
(2007,0.298495)
(2008,0.354835)
(2009,0.498079)
(2010,0.275957)
(2011,0.213544)
(2012,0.372017)
(2013,0.305522)
(2014,0.323677)
(2015,0.318475)
(2016,0.279966)
(2017,0.330230)
(2018,0.345915)
(2019,0.349438)
(2020,0.288933)
(2021,0.363689)
(2022,0.383781)
(2023,0.437357)
(2024,0.393505)
};

\addplot[color=ForestGreen, dashdotted, mark=square*, mark size=1pt, mark options={solid}] coordinates {
(1950,0.412249)
(1951,0.417447)
(1952,0.384379)
(1953,0.357969)
(1954,0.365406)
(1955,0.286789)
(1956,0.265155)
(1957,0.335652)
(1958,0.531637)
(1959,0.571699)
(1960,0.451159)
(1961,0.472511)
(1962,0.495837)
(1963,0.459829)
(1964,0.544667)
(1965,0.496164)
(1966,0.585709)
(1967,0.524951)
(1968,0.579635)
(1969,0.526793)
(1970,0.558597)
(1971,0.566770)
(1972,0.527123)
(1973,0.520264)
(1974,0.549030)
(1975,0.543013)
(1976,0.531259)
(1977,0.555015)
(1978,0.528973)
(1979,0.524650)
(1980,0.515080)
(1981,0.511417)
(1982,0.532377)
(1983,0.537642)
(1984,0.547248)
(1985,0.527576)
(1986,0.492902)
(1987,0.515595)
(1988,0.449407)
(1989,0.451598)
(1990,0.443933)
(1991,0.450616)
(1992,0.483190)
(1993,0.508524)
(1994,0.490007)
(1995,0.514865)
(1996,0.490751)
(1997,0.534939)
(1998,0.452142)
(1999,0.501304)
(2000,0.470121)
(2001,0.497353)
(2002,0.445855)
(2003,0.501970)
(2004,0.406498)
(2005,0.481106)
(2006,0.432122)
(2007,0.412883)
(2008,0.488666)
(2009,0.518740)
(2010,0.427020)
(2011,0.341479)
(2012,0.460320)
(2013,0.385928)
(2014,0.381604)
(2015,0.393716)
(2016,0.383613)
(2017,0.404163)
(2018,0.414598)
(2019,0.390629)
(2020,0.429763)
(2021,0.401808)
(2022,0.401466)
(2023,0.390070)
(2024,0.405358)
};

\draw [blue,thick,dotted] (1967,\pgfkeysvalueof{/pgfplots/ymin}) -- (1967,\pgfkeysvalueof{/pgfplots/ymax});
\draw [blue,thick,dotted] (1996,\pgfkeysvalueof{/pgfplots/ymin}) -- (1996,\pgfkeysvalueof{/pgfplots/ymax});
\draw [blue,thick,dotted] (2009,\pgfkeysvalueof{/pgfplots/ymin}) -- (2009,\pgfkeysvalueof{/pgfplots/ymax});

\draw [ForestGreen,thick,dotted] (1965,\pgfkeysvalueof{/pgfplots/ymin}) -- (1965,\pgfkeysvalueof{/pgfplots/ymax});
\end{axis}
\end{tikzpicture}
\end{subfigure}

\vspace{0.5cm}
\begin{subfigure}{\textwidth}
\caption{Weight vector: $w_k = \sqrt{1/k}$ \citep{Ausloos2024a}}
\label{Fig6b}

\begin{tikzpicture}
\begin{axis}[
width = \textwidth, 
height = 0.6\textwidth,
xlabel style = {align=center, font=\small},
xlabel = {Year},
ylabel style = {align=center, font=\small},
ylabel = {Normalised weighted distance},
xmin = 1949.5,
xmax = 2024.5,
ymajorgrids,
xtick = {1950,1960,1970,1980,1990,2000,2010,2020}, 
scaled x ticks = false,
xticklabel style = {/pgf/number format/1000 sep=},
legend style = {font=\small,at={(0.25,-0.15)},anchor=north west,legend columns=3},
legend entries = {Start--start$\qquad$, Finish--finish},
]

\addplot[blue, dashed, mark=*, mark size=1pt, mark options={solid}] coordinates {
(1950,0.151440)
(1951,0.223784)
(1952,0.278988)
(1953,0.203299)
(1954,0.246657)
(1955,0.165860)
(1956,0.229615)
(1957,0.224617)
(1958,0.334095)
(1959,0.333471)
(1960,0.395921)
(1961,0.277534)
(1962,0.319027)
(1963,0.249427)
(1964,0.213956)
(1965,0.321357)
(1966,0.298658)
(1967,0.310415)
(1968,0.395818)
(1969,0.330548)
(1970,0.354380)
(1971,0.346477)
(1972,0.307097)
(1973,0.316202)
(1974,0.328505)
(1975,0.323967)
(1976,0.341768)
(1977,0.302537)
(1978,0.259541)
(1979,0.325621)
(1980,0.344733)
(1981,0.279075)
(1982,0.260245)
(1983,0.277887)
(1984,0.275610)
(1985,0.280034)
(1986,0.216995)
(1987,0.183602)
(1988,0.213077)
(1989,0.239920)
(1990,0.214054)
(1991,0.197211)
(1992,0.226513)
(1993,0.214830)
(1994,0.233647)
(1995,0.164710)
(1996,0.170894)
(1997,0.269234)
(1998,0.211418)
(1999,0.249076)
(2000,0.250700)
(2001,0.208777)
(2002,0.216593)
(2003,0.337011)
(2004,0.326007)
(2005,0.365255)
(2006,0.340152)
(2007,0.255132)
(2008,0.296853)
(2009,0.429284)
(2010,0.224565)
(2011,0.183950)
(2012,0.295814)
(2013,0.251197)
(2014,0.286673)
(2015,0.295313)
(2016,0.247859)
(2017,0.294314)
(2018,0.310163)
(2019,0.305550)
(2020,0.252803)
(2021,0.313565)
(2022,0.340753)
(2023,0.383770)
(2024,0.339092)
};

\addplot[color=ForestGreen, dashdotted, mark=square*, mark size=1pt, mark options={solid}] coordinates {
(1950,0.371717)
(1951,0.372842)
(1952,0.376649)
(1953,0.307413)
(1954,0.335324)
(1955,0.241073)
(1956,0.220130)
(1957,0.292399)
(1958,0.476033)
(1959,0.509606)
(1960,0.407583)
(1961,0.404130)
(1962,0.439191)
(1963,0.420797)
(1964,0.486874)
(1965,0.460029)
(1966,0.533730)
(1967,0.472753)
(1968,0.523427)
(1969,0.474547)
(1970,0.497951)
(1971,0.515238)
(1972,0.474736)
(1973,0.468357)
(1974,0.483381)
(1975,0.480953)
(1976,0.474361)
(1977,0.494352)
(1978,0.463109)
(1979,0.457853)
(1980,0.458321)
(1981,0.437699)
(1982,0.460551)
(1983,0.473842)
(1984,0.494389)
(1985,0.465134)
(1986,0.444879)
(1987,0.462876)
(1988,0.410695)
(1989,0.374527)
(1990,0.376921)
(1991,0.391679)
(1992,0.439823)
(1993,0.468490)
(1994,0.441848)
(1995,0.466465)
(1996,0.448145)
(1997,0.479754)
(1998,0.411527)
(1999,0.454176)
(2000,0.438827)
(2001,0.460276)
(2002,0.422439)
(2003,0.451830)
(2004,0.373879)
(2005,0.433215)
(2006,0.394619)
(2007,0.373428)
(2008,0.425389)
(2009,0.460121)
(2010,0.372485)
(2011,0.312077)
(2012,0.396424)
(2013,0.348335)
(2014,0.346178)
(2015,0.369050)
(2016,0.351714)
(2017,0.369195)
(2018,0.382680)
(2019,0.357814)
(2020,0.392168)
(2021,0.359893)
(2022,0.371565)
(2023,0.369723)
(2024,0.350159)
};

\draw [blue,thick,dotted] (1967,\pgfkeysvalueof{/pgfplots/ymin}) -- (1967,\pgfkeysvalueof{/pgfplots/ymax});
\draw [blue,thick,dotted] (1996,\pgfkeysvalueof{/pgfplots/ymin}) -- (1996,\pgfkeysvalueof{/pgfplots/ymax});
\draw [blue,thick,dotted] (2009,\pgfkeysvalueof{/pgfplots/ymin}) -- (2009,\pgfkeysvalueof{/pgfplots/ymax});

\draw [ForestGreen,thick,dotted] (1965,\pgfkeysvalueof{/pgfplots/ymin}) -- (1965,\pgfkeysvalueof{/pgfplots/ymax});
\end{axis}
\end{tikzpicture}
\end{subfigure}

\caption{Normalised weighted distance of start--start \\ and finish--finish rankings with structural breaks}
\label{Fig7}
\end{figure}


Analogously, according to the weighted distances in Figure~\ref{Fig7}, the start rankings have become increasingly competitive relative to the finish rankings. The break points are almost identical to those in Figure~\ref{Fig5}. There are again three changes in 1967, 1996, 2009 for start-start positions, and one change in 1965 for finish-finish positions. Last but not least, the top-heavy weights $w_k = 1/k$ indicate a more intense competition.

Interestingly, as revealed by all measures, competitive balance based on start positions is moving closer and closer to competitive balance based on finish positions after a substantial gap between 1965 and 1995. In the last seasons, the rankings in the qualifications and the races are fully comparable in terms of their volatility.

\section{Conclusions} \label{Sec5}

This paper has developed an approach to measuring competitive balance in Formula One by using weighted distances between race rankings. Building on the ranking-based method of \citet{PeetersWesselbaum2023}, we argue that the standard Kem\'eny distance can be refined by taking into account the position-dependent importance of swaps. In particular, a change at the top of the ranking is likely to be more relevant to both decision-makers and spectators compared to a change at the bottom of the ranking. The weighted distance suggested by \citet{Can2014} can address this issue.

Using the longest available Formula One dataset, covering all seasons from 1950 to 2024, we have calculated both standard and weighted competitive balance indicators at the level of drivers.
The key results emerge:
(1) the long-run evolution of competitive balance is unexpectedly robust with respect to the weighting scheme;
(2) the intensity of competition is higher among the fastest drivers; and
(3) competitive balance is more unfavourable in the last two decades than ever before. 

According to the structural break analysis, competitive balance in Formula One has evolved through distinct regimes. Most shifts detected by the Bai--Perron procedure can be interpreted in light of major technical, regulatory, and institutional developments in the sport. Therefore, competitive balance is closely connected to the regulatory environment of Formula One, and decision-makers have powerful tools to influence the level of competitiveness. 

Our analysis highlights the importance of distinguishing between different dimensions of competitive balance. For example, the trends in the start--start and finish--finish rankings do not always move identically, reinforcing the complex nature of competitive balance in racing sports. 

In the proposed methodology, the choice of weights remains an important issue to explore. Future research could estimate position weights empirically, for example, from fan surveys, official point-scoring systems, or television viewing data. Such an approach would help determine the relative importance of changes in the top and bottom positions. 

Finally, the framework developed here can be extended beyond Formula One. Other racing competitions, such as IndyCar, MotoGP, NASCAR,  cycling stage races, or endurance racing, also generate full rankings and involve position-dependent fan interest. Weighted distances of rankings provide a flexible tool for comparing competitive balance in all these sports. Applying them to sports where rankings play a central role may contribute to a more nuanced understanding of outcome uncertainty.

\section*{AI use}
\addcontentsline{toc}{section}{AI use}

We used Microsoft Copilot to improve the linguistic quality of the manuscript and to get ideas for the illustrative examples. The authors reviewed and edited the content as needed and take full responsibility.

\section*{Acknowledgements}
\addcontentsline{toc}{section}{Acknowledgements}

The research was supported by the National Research, Development and Innovation Office under Grants Advanced 152220, FK 145838 and PD 153835, and the J\'anos Bolyai Research Scholarship of the Hungarian Academy of Sciences.

\bibliographystyle{apalike}
\bibliography{All_references}

\clearpage
\setcounter{figure}{0}
\renewcommand{\thefigure}{A.\arabic{figure}}

\setcounter{table}{0}
\renewcommand{\thetable}{A.\arabic{table}}

\section*{Appendix}
\addcontentsline{toc}{section}{Appendix}

\begin{table}[!ht]
  \centering
  \caption{The number of drivers and races in Formula One seasons}
  \label{Table_A1}
  \rowcolors{0}{}{gray!20}
    \begin{tabularx}{0.8\textwidth}{LCCLCC} \toprule
    Season & Drivers & Races & Season & Drivers & Races \\ \bottomrule
    1950  & 81    & 7     & 1988  & 36    & 16 \\
    1951  & 84    & 8     & 1989  & 47    & 16 \\
    1952  & 105   & 8     & 1990  & 40    & 16 \\
    1953  & 108   & 9     & 1991  & 41    & 16 \\
    1954  & 97    & 9     & 1992  & 37    & 16 \\
    1955  & 84    & 7     & 1993  & 35    & 16 \\
    1956  & 85    & 8     & 1994  & 46    & 16 \\
    1957  & 76    & 8     & 1995  & 35    & 17 \\
    1958  & 87    & 11    & 1996  & 24    & 16 \\
    1959  & 88    & 9     & 1997  & 28    & 17 \\
    1960  & 91    & 10    & 1998  & 23    & 16 \\
    1961  & 62    & 8     & 1999  & 24    & 16 \\
    1962  & 61    & 9     & 2000  & 23    & 17 \\
    1963  & 62    & 10    & 2001  & 26    & 17 \\
    1964  & 41    & 10    & 2002  & 23    & 17 \\
    1965  & 54    & 10    & 2003  & 24    & 16 \\
    1966  & 33    & 9     & 2004  & 25    & 18 \\
    1967  & 45    & 11    & 2005  & 27    & 19 \\
    1968  & 43    & 12    & 2006  & 27    & 18 \\
    1969  & 31    & 11    & 2007  & 26    & 17 \\
    1970  & 43    & 13    & 2008  & 22    & 18 \\
    1971  & 50    & 11    & 2009  & 25    & 17 \\
    1972  & 42    & 12    & 2010  & 27    & 19 \\
    1973  & 43    & 15    & 2011  & 28    & 19 \\
    1974  & 62    & 15    & 2012  & 25    & 20 \\
    1975  & 52    & 14    & 2013  & 23    & 19 \\
    1976  & 54    & 16    & 2014  & 24    & 19 \\
    1977  & 61    & 17    & 2015  & 22    & 19 \\
    1978  & 46    & 16    & 2016  & 24    & 21 \\
    1979  & 36    & 15    & 2017  & 25    & 20 \\
    1980  & 41    & 14    & 2018  & 20    & 21 \\
    1981  & 39    & 15    & 2019  & 20    & 21 \\
    1982  & 40    & 16    & 2020  & 23    & 17 \\
    1983  & 35    & 15    & 2021  & 21    & 22 \\
    1984  & 35    & 16    & 2022  & 22    & 22 \\
    1985  & 36    & 16    & 2023  & 22    & 22 \\
    1986  & 32    & 16    & 2024  & 24    & 24 \\
    1987  & 32    & 16    &       &       &  \\ \bottomrule
    \end{tabularx}
\end{table}

\end{document}